%% file: paper.tex
\documentclass[acmsmall, screen]{acmart}
\acmJournal{PACMPL}
\acmVolume{1}
\acmNumber{CONF} 
\acmArticle{1}
\acmYear{2018}
\acmMonth{1}
\acmDOI{} 
\startPage{1}

\setcopyright{none}
\renewcommand\footnotetextcopyrightpermission[1]{}
\usepackage{amssymb}
\usepackage{pifont}

\newcommand{\cmark}{\ding{51}}%
\newcommand{\xmark}{\ding{55}}%

\usepackage[normalem]{ulem}
\newcommand*{\tool}{HardTaint }

\usepackage{tikz}
\usepackage{amsmath}
\usepackage{wrapfig}

\usepackage{multirow}
\usepackage{MnSymbol}

\usepackage{filecontents}
\usepackage{float}
\usepackage{subfig}
\usepackage{enumitem}

\usepackage{xcolor}
\usepackage{bbding}
\usepackage{cancel}

\usepackage{algorithmic}
\usepackage[ruled, vlined, linesnumbered]{algorithm2e}

\usepackage{graphicx}

\newcommand{\pie}[1]{%
\begin{tikzpicture}
 \draw (0,0) circle (0.8ex);
 \fill (0.8ex,0) arc (0:#1:0.8ex) -- (0,0) -- cycle;
\end{tikzpicture}%
}

\newcommand{\dashedbox}[1]{%
  \tikz[baseline=(text.base)]{
    \node[draw, dashed, inner sep=2pt, outer sep=0pt] (text) {#1};
  }%
}

\newcommand{\solidbox}[1]{%
  \tikz[baseline=(text.base)]{
    \node[draw, inner sep=2pt, outer sep=0pt] (text) {#1};
  }%
}

\newcommand{\zyy}[1]{\textcolor{red}{{#1}}}

\newcommand{\eg}{\hbox{\emph{e.g.}}\xspace}
\newcommand{\ie}{\hbox{\emph{i.e.}}\xspace}

\newcommand{\MyPara}[1]{\vspace{1pt}\noindent\textbf{\textit{#1}}\hspace{0.2em}}

\begin{document}

\title{HardTaint: Production-Run Dynamic Taint Analysis via Selective Hardware Tracing}

\author{Yiyu Zhang}
\authornote{Also with State Key Laboratory for Novel Software Technology at Nanjing University.}
\affiliation{%
  \institution{Nanjing University}
  \country{China}
  }
\email{zhangyy0721@smail.nju.edu.cn}

\author{Tianyi Liu}
\authornotemark[1]
\affiliation{%
  \institution{Nanjing University}
  \country{China}
  }
\email{tianyiliu@smail.nju.edu.cn}

\author{Yueyang Wang}
\authornotemark[1]
\affiliation{%
  \institution{Nanjing University}
  \country{China}
  }
\email{502022330054@smail.nju.edu.cn}

\author{Yun Qi}
\authornotemark[1]
\affiliation{%
  \institution{Nanjing University}
  \country{China}
  }
\email{qiyun@smail.nju.edu.cn}

\author{Kai Ji}
\authornotemark[1]
\affiliation{%
  \institution{Nanjing University}
  \country{China}
  }
\email{jikai@smail.nju.edu.cn}

\author{Jian Tang}
\authornotemark[1]
\affiliation{%
  \institution{Nanjing University}
  \country{China}
  }
\email{tangjian@smail.nju.edu.cn}

\author{Xiaoliang Wang}
\authornotemark[1]
\affiliation{%
  \institution{Nanjing University}
  \country{China}
  }
\email{waxili@nju.edu.cn}

\author{Xuandong Li}
\authornotemark[1]
\affiliation{%
  \institution{Nanjing University}
  \country{China}
  }
\email{lxd@nju.edu.cn}

\author{Zhiqiang Zuo}
\authornotemark[1]
\authornote{Corresponding author.}
\affiliation{%
  \institution{Nanjing University}
  \country{China}
  }
\email{zqzuo@nju.edu.cn}

\begin{abstract}

\input{abstract.tex}

\end{abstract}
\maketitle

\section{Introduction}
\label{sec:intro}
\input{introduction.tex}

\section{Background}
\label{sec:background}
\input{background}

\section{Overview}
\label{sec:overview}
\input{overview}

\section{Runtime Selective Tracing}
\label{sec:online-selective-tracing}
\input{online-selective-tracing}

\section{Real-time Parallel Analysis}
\label{sec:offline-taint-graph-parallel-analysis}
\input{offline-taint-graph-parallel-analysis}

\section{Implementation}
\label{sec:implementation}
\input{implementation}

\section{Evaluation}
\label{sec:evaluation}
\input{evaluation}

\section{Discussion}
\label{sec:discussion}
\input{discussion.tex}

\section{Related Work}
\label{sec:related-work}
\input{related-work}

\section{Conclusion}
\label{sec:conclusion}
\input{conclusion}


\bibliographystyle{plain}
\bibliography{paper}

\end{document}

%% file: abstract.tex
Dynamic taint analysis (DTA), as a fundamental analysis technique, is widely used in security, privacy, and diagnosis, etc.
As DTA demands to collect and analyze massive taint data online, it suffers extremely high runtime overhead. 
Over the past decades, numerous attempts have been made to lower the overhead of DTA. 
Unfortunately, the reductions they achieved are marginal, causing DTA only applicable to the debugging/testing scenarios. 
In this paper, we propose and implement HardTaint, a system that can realize production-run dynamic taint tracking. 
HardTaint adopts a hybrid and systematic design which combines static analysis, selective hardware tracing and parallel graph processing techniques. 
The comprehensive evaluations demonstrate that HardTaint introduces only around 9\% runtime overhead which is an order of magnitude lower than the state-of-the-arts, while without sacrificing any taint detection capability. 


%% file: introduction.tex
Dynamic taint analysis (DTA for short) as a fundamental program analysis technique is widely adopted in a vast variety of application areas, including security, privacy, and diagnostic \cite{newsome2005dynamic, suh2004secure,sharif2009automatic, yin2007panorama,zhu2011tainteraser,wang2010taintscope, attariyan2010automating, enck2014taintdroid}, etc. 
Compared to static taint analysis \cite{arzt2014flowdroid, tripp2009taj, sridharan2011f4f, grech2017p} which suffers from high false positives, DTA has the advantage of high precision.
However, as DTA demands to collect and analyze massive dynamic taint data online, it dramatically slows down the program execution.  
As reported \cite{kemerlis2012libdft, clause2007dytan}, the runtime overhead suffered by DTA readily reaches 400\%-3000\%.
Such high runtime overhead relegates DTA to pre-deployment testing only, which can only detect taint behaviors under the limited test suite.
Unfortunately, despite extensive in-house testing, taint problems commonly escape to production runs, usually leading to silent privacy leakage or subtle memory corruptions. 
Beyond question, an ultra-low overhead DTA  which is able to real-time monitor taint behaviors in production-run environments is greatly desirable \cite{enck2014taintdroid,qin2006lift,ouyang2023mirrortaint}.


\MyPara{State of the Arts.} 
Over the past decades, numerous attempts have been made to lower the runtime overhead of DTA by reducing unnecessary taint tracking via program analysis \cite{zhu2011tainteraser, jee2012general, banerjee2019iodine, chen2021selectivetaint}, generating fast paths \cite{qin2006lift, davanian2019decaf++, galea2020taint}, and offloading taint logic from the original execution \cite{cui2015practical, jee2013shadowreplica, ming2015taintpipe, ming2016straighttaint}. 
Unfortunately, all these approaches inevitably have to conduct expensive online taint propagation and/or intensive runtime information collection. As a result, their overheads unsurprisingly stay 172\%-310\% on the common SPEC CPU benchmarks as reported \cite{jee2013shadowreplica,ming2015taintpipe,ming2016straighttaint,jee2012general}, 
which is apparently far from satisfactory and unacceptable in the production-run environment.
Another line of work accelerates taint analysis via dedicated hardware instructions, registers, and co-processors \cite{bosman2011minemu,pilato2018tainthls,kannan2009decoupling, chen2008flexible}. Although promising results are shown through simulation experiments, they can hardly be practical without the need for extensive hardware redesign.

\MyPara{Our Work.}
In this paper, we develop HardTaint, a system that can realize on-the-fly dynamic taint tracking with ultra-low overhead on commodity hardware. 
We noticed that modern hardware tracing modules widely integrated with the commodity CPUs (\emph{e.g.}, Intel PT \cite{intelpt}, ARM ETM \cite{armetm}) provides capabilities for precise program tracing with minimal overhead. 
As such, we revisit the design of decoupled DTA \cite{cui2015practical, jee2013shadowreplica, ming2015taintpipe, ming2016straighttaint} and leverage Intel PT to implement it.
As a decoupled design,  HardTaint consists of two separate steps, namely runtime information collection and real-time taint propagation. 
The former collects necessary control and data information at runtime by leveraging hardware tracing (in particular Intel PT \cite{intelpt}). 
Meanwhile, the later constantly takes as input the runtime data recorded and performs taint tracking at real-time.  
%
While the idea appears simple at a high level, doing so efficiently (\emph{i.e.}, the collection and analysis part should have sufficiently low overhead and latency, respectively) and precisely (\emph{i.e.}, the taint results should be as precise as the usual DTA) requires overcoming three major challenges.

First, naively tracing all the taint information via hardware still yields prohibitively high overhead in our experiments (see \S\ref{sec:optimizations}). 
This is because taint propagation demands a rich vein of runtime information, including not only control-flow but also plentiful data values. 
Despite that Intel PT can handle control-flow information with negligible overhead (2\%-5\%~\cite{sharma2016hardware,ptforvm-2017}), a massive number of expensive instructions (\ie, \emph{ptwrite}) have to be instrumented and executed to record the intensive data values needed by taint tracking. 

Second, naively recording the intensive taint information via hardware results in severe data loss, causing imprecision of taint tracking. 
As CPU can execute instructions much faster than memory can keep up, 
the speed of trace generation exceeds trace dumping.  
As a result, certain trace data would be lost especially if the runtime information to be recorded is too dense. 
Our empirical studies demonstrate that naive hardware tracing causes frequent data loss, which ultimately fails the taint tracking (see \S\ref{sec:optimizations}). 


Third, performing taint analysis on the massive-scale trace data is costly and inefficient. 
The analysis process needs to load hardware trace data into memory, decodes it to recover runtime information, and performs taint propagation to produce the final report.    
It takes considerably long time, which is usually 6.18x longer than the program's execution time in our experiments (see \S \ref{sec:optimizations}). 
Such inefficiency turns into the unacceptable latency between taint triggering and reporting.

To address the above challenges, \tool adopts a hybrid and systematic design which combines static analysis, selective hardware tracing, pipelining, and parallel graph processing techniques. 
In particular, we devise static binary analysis and selective hardware tracing to tackle the first two challenges. 
The underlying reason for both high overhead and data loss of naive hardware tracing is that too intensive control (\ie, branch targets) and data (\ie, registers)  values need to be recorded at runtime.  
State-of-the-art static analysis-based approaches such as SelectiveTaint \cite{chen2021selectivetaint} can be exploited to eliminate the branches and registers which are irrelevant to taint tracking. 
Unfortunately, its effectiveness is marginal shown by our experimental study in \S\ref{sec:optimizations} -- there are still a prohibitively large number ($ \sim$70\% on average) of them required to be recorded at runtime. The reason is twofold: 1) the static analysis adopted is conservative; more importantly 2) branches and registers actually involved in taint tracking are still too intensive.
In fact, many branches and registers even involved in taint propagation are not necessarily recorded at runtime. 
Thanks to the existence of intrinsic static code information, the values of some branches and registers can be statically deduced from that of others.
Therefore, we can identify a small essential set of branches and registers, and only record their values at runtime, while all others can be statically derived.
Based on this insight, \tool employs a dedicated static analysis to identify a minimal set of necessary control (\ie, branch targets) and data (\ie, registers) to be traced (see \S\ref{sec:identification-analysis}), and adopts selective hardware tracing to only capture such essential information at runtime (see \S\ref{sec:binary-rewriter}). 
In this way, the online part of \tool not only has adequately low overhead, but also produces the essential runtime data for the subsequent taint analysis. 
Our experimental evaluations show that the percentage of instructions to be traced is decreased from 70\% to 11\%, runtime overhead from 79\% to 9\% on average (see \S\ref{sec:optimizations}). 

To address the third challenge, existing work such as FlowMatrix \cite{ji2022flowmatrix} proposes a matrix-based representation and enables GPUs to accelerate offline taint propagation. 
However, it cannot be simply employed in our scenario since not all of the control and data values involved in taint propagation can be directly read from traces. A great portion of values need to be deduced statically. Such deduction process could be the bottleneck.
Moreover, the acceleration by FlowMatrix is limited, only 5.6x speedup over the CPU-based baseline as reported in \cite{ji2022flowmatrix}. 
Here, we propose a systematic solution that integrates analysis offloading, pipelining, and tailored parallel analysis performing static deduction and taint propagation simultaneously. 
In particular, 
1) we offload the taint analysis to a separate analysis machine with trace transferred via the high-speed Remote Direct Memory Access (RDMA) \cite{rdma-wiki}, thus avoiding the resource preemption and performance degradation on the production machine; 2) we let trace generation, trace transferring \& decoding, and taint propagation execute as a pipeline so as to shorten the end-to-end analysis time; 3) we propose highly parallel decoding (\S \ref{sec:decoder}) and taint propagation (\S\ref{sec:taint-analysis}) which offer in-time analysis capability over massive data. Put them all together, \tool ultimately achieves super-low analysis latency.
The average speedup of \tool over the CPU-based baseline reaches 100x in our experiments (see \S\ref{sec:optimizations}).

\begin{table}[htb!]
\caption{Comparison of HardTaint with other approaches.}\vspace{-0.5em}
\label{tab:intro-comparison}
\centering
\scalebox{0.85}{
\begin{tabular}{l|c|c|c}
\hline
Approach &
  \begin{tabular}[c]{@{}c@{}}low\\ overhead\end{tabular} &
  \begin{tabular}[c]{@{}c@{}}high\\ precision\end{tabular} &
  \begin{tabular}[c]{@{}c@{}}low\\ latency\end{tabular}  \\ \hline \hline
Software-based \cite{zhu2011tainteraser, chen2021selectivetaint, qin2006lift, galea2020taint} & \textcolor{red}{\xmark\xmark\xmark}       & \textcolor{green}{\cmark} & \textcolor{green}{\cmark} \\
Software-decoupled \cite{cui2015practical, jee2013shadowreplica, ming2015taintpipe, ming2016straighttaint} & \textcolor{red}{\xmark\xmark}          & \textcolor{green}{\cmark} & \textcolor{red}{\xmark}     \\
Naive hardware tracing (NHT in \S \ref{sec:optimizations}) &
  \textcolor{red}{\xmark} &
  \textcolor{red}{\xmark} &
  \textcolor{red}{\xmark} \\
HardTaint (this work)       & \textcolor{green}{\cmark} & \textcolor{green}{\cmark} & \textcolor{green}{\cmark} \\ \hline
\end{tabular}
}
\end{table}

\MyPara{Results.}
We implemented \tool and evaluated it over a comprehensive set of well-known open-sourced subjects, including Unix utilities, SPEC CPU INT 2006, three network daemon programs, and two read-world large-scale applications (\ie, PHP and MySQL).
The experimental results show that the runtime overhead introduced by \tool is only around 9.06\% on average, which is orders of magnitude lower than the state-of-the-art approaches. 
The static identification analysis is both efficient and scalable. It only takes a few minutes on average for moderate-sized programs and scales well to millions lines of code. 
Moreover, the parallel taint propagation has pretty high scalability -- the analysis latency of HardTaint is only 0.58 seconds in general. 
To validate the effectiveness of HardTaint, we chose  8 commonly used CVEs and reproduced each of them. The experimental results indicate that \tool is able to detect all these vulnerabilities.
HardTaint's artifacts are available on Github (link removed for anonymity).  
A qualitative comparison between \tool and the representative existing approaches can be found in Table \ref{tab:intro-comparison}.

In summary, we make the following contributions.
\begin{itemize}
   \item We develop the first system to the best of our knowledge leveraging commodity hardware tracing (\ie, Intel PT) to realize practical dynamic taint analysis for production use.
   \item We devise the novel and dedicated static identification analysis to lower runtime overhead, and parallel taint propagation to shorten the latency of reporting.
   \item We implement a prototype called HardTaint, and conduct the evaluations over a comprehensive set of well-known open-sourced subjects to validate HardTaint's low runtime overhead, low latency, and high precision.
\end{itemize}

\MyPara{Outline.}
The rest of the paper is organized as follows. \S \ref{sec:background} gives the necessary background of dynamic taint analysis and Intel processor tracing. 
\S \ref{sec:overview} provides the overview of HardTaint. 
\S \ref{sec:online-selective-tracing} and \S \ref{sec:offline-taint-graph-parallel-analysis} describe the key components we proposed, followed by the implementation of \tool in \S \ref{sec:implementation}.  
We present the empirical evaluations in \S \ref{sec:evaluation}. 
Certain issues 
are discussed in \S \ref{sec:discussion}.
We talk about the related work in \S \ref{sec:related-work}. Finally, \S \ref{sec:conclusion} concludes.

%% file: background.tex
\subsection{Dynamic Taint Analysis (DTA)}
DTA typically consists of three key components: taint sources, taint sinks, and taint propagation.
DTA maintains taint status for registers and memory addresses.

\MyPara{Taint Sources and Sinks.} 
Both taint sources and sinks are the essential inputs of DTA.
Taint sources are program points or memory locations where data of interest is introduced. 
Generally, data of interest corresponds to specific variables and memory offsets, which are returned from a specific function, or read from I/O stream (\eg, file system, network, or keyboard). 
Taint sinks are program points or memory locations where DTA checks the taint status against the particular security applications (\eg, check sensitive parameters for information leakage or check jump targets for control flow hijacks).

\MyPara{Taint Propagation.} 
Given the sources and sinks, DTA propagates taint status among instruction operands according to the instruction semantics during program execution.
For instance, given an instruction \emph{<and eax, ebx>}, a taint propagation rule for instruction \emph{and} prescribes that the new status of \emph{eax} is updated as a bit-wise or of status of \emph{eax} and \emph{ebx}. 
In addition to the explicit taint propagation on data dependence, there also exists implicit taint propagation on control dependence.
Given an instruction $inst$, its destination operands are tainted if the source operands of a branch $br$ are tainted where $inst$ is control dependent on $br$.

\subsection{Intel Processor Tracing (PT)}
Intel PT is a new hardware feature provided by Intel processors to implement instruction-level control-flow and data tracing. 
PT can record control-flow information such as instruction pointers, indirect branch targets, and directions of conditional branches with extremely low overhead.
These information are carefully encoded into various types of trace packets, including Packet Generation Enable (PGE), Packet Generation Disable (PGD), Taken Not-Taken (TNT), and Target Instruction Pointer (TIP), Flow Update Packet (FUP), etc. 
Apart from the control-flow information, PT can also collect data values via \emph{ptwrite} instruction \cite{intelpt}. The data are recorded as PTW packets.
To minimize the impact on performance, PT directly writes trace packets into physical memory buffer bypassing caches and TLB.
Since the trace data collected is highly compressed, a decoding process is required to recover the execution information from the trace packets.
In HardTaint, we exploit Intel PT to acquire both control and data information needed by DTA. Differently, we propose dedicated design to achieve selective tracing, and accelerate the decoding process via parallelism.

%% file: overview.tex
\tool aims to provide practical dynamic taint tracking within a production environment. 
We envision a production environment similar to Google’s real-world data center, where the applications under analysis running on a production machine generate traces at runtime, and dedicated analysis machines within the data center process these traces simultaneously \cite{google-profiling,prorace-2017, wesolowski2021datacenter}.
The production and analysis machines are connected via high-speed RDMA network cards, which nowadays is the standard equipment in modern data centers \cite{rdma-datacenters, rdma-for-enterprise}. 

\vspace{-1em}
\begin{figure*}[htbp!]
	\centering
	\includegraphics[scale=0.48]{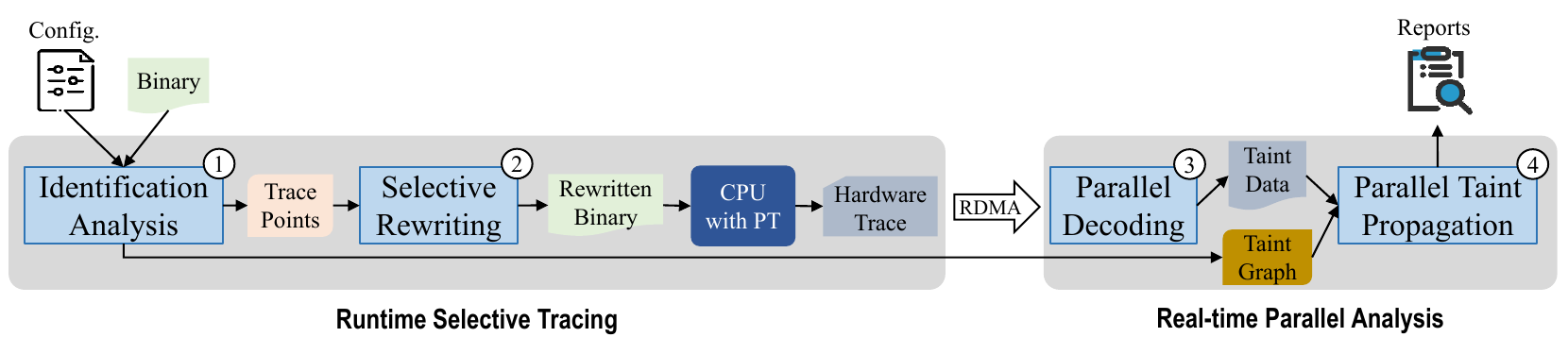}
  \vspace{-0.5em}
	\caption{Workflow of HardTaint. \label{fig:overview}}
\end{figure*}

Figure \ref{fig:overview} gives the workflow of HardTaint.
Basically, it consists of two phases: runtime selective tracing and real-time parallel analysis.  
In the tracing phase, \tool leverages static identification analysis and selective binary rewriting to minimize the runtime information recorded as much as possible, thus achieving ultra-low overhead. 
Meanwhile, the analysis phase constantly takes as input the hardware trace generated  and performs parallel taint propagation to produce in-time report.  To avoid mutual interference, the two phases are conducted on two separate machines, between which the traces are transferred via the high-speed RDMA \cite{rdma-wiki}. 

\vspace{-0.5em}
\begin{figure}[htb!]
	\centering
	\includegraphics[scale=0.4]{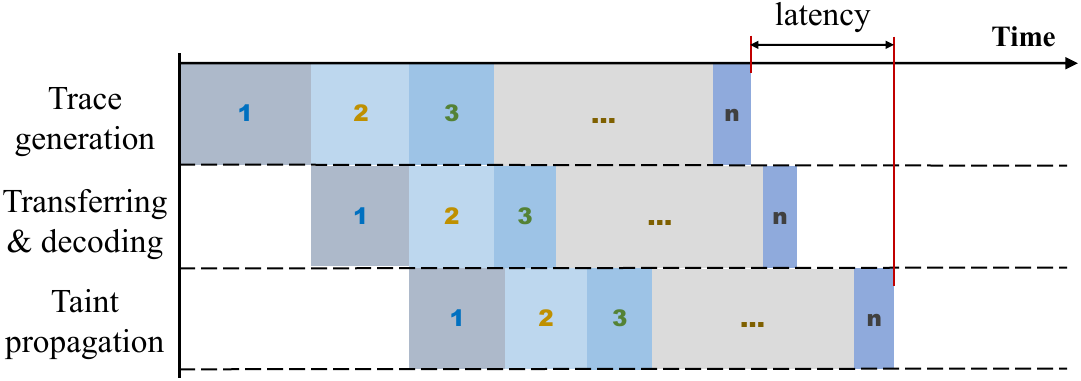}
  \vspace{-0.5em}
	\caption{Pipeline of HardTaint. \label{fig:pipeline}}
\end{figure}

Specifically, \tool takes the binary under analysis and user-defined taint configurations as input. It first applies static identification analysis (cf. \ding{172}) to pinpoint a minimal set of necessary trace points, each of which corresponds to a branch target or a register value. 
Given the essential trace points identified, a binary rewriter (cf. \ding{173}) inserts the corresponding instructions into the original binary so as to fulfill selective hardware tracing. 
After that, the rewritten binary is executed on a production machine with Intel PT enabled. 
Along with the execution, hardware traces are generated constantly.  
Meanwhile, another analysis machine repeatedly reads the hardware traces via RDMA, invokes a parallel decoder (cf. \ding{174}) to recover the runtime information needed, and performs parallel taint propagation (cf. \ding{175}) to produce taint results. 
Note that the three steps namely trace generation, trace  transferring  \& decoding\footnote{As the transferring time via RDMA is negligible compared to that of decoding, we merge transferring with decoding into one step in the pipeline.}, and taint propagation, are executed as a pipeline illustrated as Figure \ref{fig:pipeline}, so that the taint results can be obtained as fast as possible.




\begin{figure}[htb!]
\subfloat[][original binary]{\hspace{0em}\label{fig:overview-original-binary}\includegraphics[width=0.17\textwidth,height=0.27\textwidth]{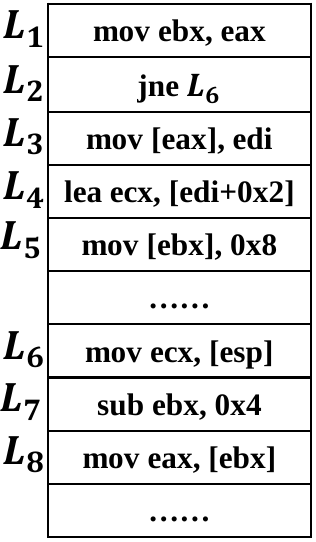}}
\hspace{2em}
    \subfloat[][naive trace points]{\label{fig:overview-naive-traced-target-info}\includegraphics[width=0.15\textwidth,height=0.17\textwidth]{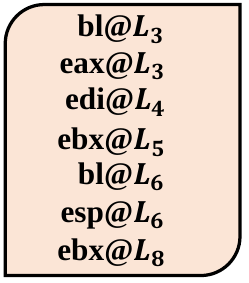}} 
    \hspace{2em}
\begin{minipage}[b][0.3\textwidth][s]{0.25\textwidth}
    \subfloat[][trace points of SelectiveTaint]{\label{fig:overview-selectivetatint-traced-target-info}\includegraphics[width=0.6\textwidth,height=0.5\textwidth]{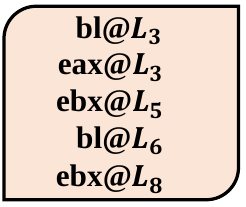}} \vfill
    \subfloat[][trace points of HardTaint]{\label{fig:overview-traced-target-info}\includegraphics[width=0.6\textwidth,height=0.3\textwidth]{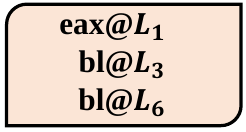}}
\end{minipage}
    \subfloat[][rewritten binary]{\label{fig:overview-rewritten-binary}\includegraphics[width=0.18\textwidth,height=0.32\textwidth]{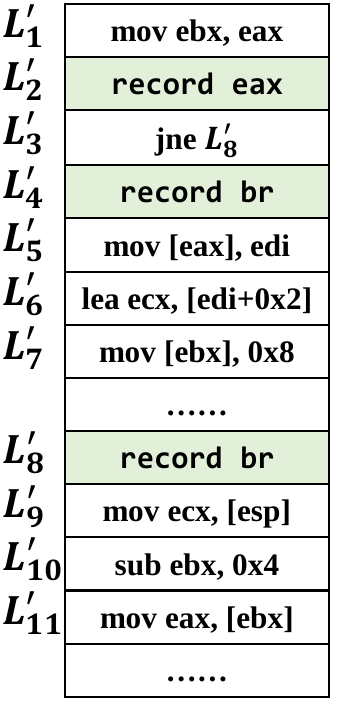}} 
\hspace{2em}
\subfloat[][taint data]{\hspace{0pt}\label{fig:overview-target-data}\includegraphics[width=0.12\textwidth,height=0.08\textwidth]{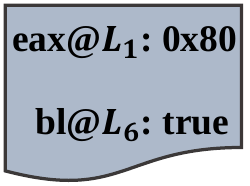}} 
    \hspace{3em}
    \subfloat[][static taint graph]{\label{fig:overview-taint-graph}\includegraphics[scale=0.43]{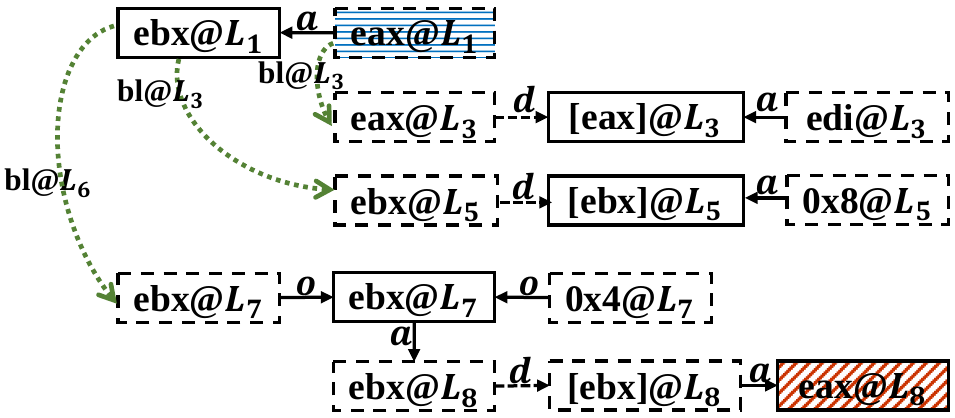}} 
    \caption{An illustrative toy example. \label{fig:overview-online-example}}
   \vspace{-0.5em}
\end{figure}




    
    

\MyPara{Example.} 
Here we use a toy example to illustrate how \tool works. 
Given a snippet of instructions under analysis shown as Figure \ref{fig:overview-original-binary}. 
Assume that the taint source and sink are $eax@L_1$ and $eax@L_8$, respectively. The goal is to determine whether the source can be propagated to the sink along the execution. To this end, we need to record all the relevant control and data information at runtime, and then replay the taint propagation to obtain the results.
Note that it is unnecessary to record the top-level register values (\eg, $eax@L_1$, $ebx@L_1$, and $ebx@L_7$), since taint between them can be propagated directly via explicit assignment \cite{ming2016straighttaint, cui2015practical}. 
However, all the register values representing memory addresses need to be captured so as to determine the propagation among memory regions. 
As a result, one naive approach is to record the branch targets $bl@L_3$ and $bl@L_6$, together with memory registers $eax@L_3$, $edi@L_4$, $ebx@L_5$, $esp@L_6$, and $ebx@L_8$, as shown in Figure \ref{fig:overview-naive-traced-target-info}. 
We can apply the state-of-the-art static analysis proposed by SelectiveTaint \cite{chen2021selectivetaint} to prune away instructions which must be unrelated to taint propagation. 
Figure \ref{fig:overview-selectivetatint-traced-target-info} gives the result after performing SelectiveTaint where $edi@L_4$ and $esp@L_6$ are safely eliminated from tracing. 
However, such result still contains a considerable amount of redundancies.
In fact, the memory registers (\ie, $eax@L_3$, $ebx@L_5$, and $ebx@L_8$) can be deduced statically according to the value of $eax@L_1$ based on the static code information. 
Based on the insight, \tool employs a dedicated static analysis to identify all the redundant registers and blocks, and pinpoints a minimal set of essential ones shown as Figure \ref{fig:overview-traced-target-info}.
It contains three points: the value of register $eax@L_1$, the two potential branch targets of $jne$, \ie, $bl@L_3$ and $bl@L_6$.   
We will elaborate the analysis in \S\ref{sec:identification-analysis}.
Next, \tool rewrites the original binary by inserting \emph{record} instructions so as to selectively trace the necessary data. 
Figure \ref{fig:overview-rewritten-binary} shows the rewritten binary code where the \emph{record} instruction logically demonstrates our intention. 
In practice, each \emph{record} is replaced with a physical \emph{ptwrite} instruction. The details will be discussed shortly in \S\ref{sec:binary-rewriter}.   
When the rewritten binary is executed on a CPU with PT, hardware trace data is generated in a compressed format. We do not show it here as it is not human-readable.  
Meanwhile, a decoder reads the trace data via RDMA and recovers the runtime values of trace points of interest (discussed in \S\ref{sec:decoder}). 
Figure \ref{fig:overview-target-data} gives the results, which reveals that the value of $eax@L_{1}$ is $0x80$, and the branch target is $L_{6}$. 
Finally, the taint propagation is performed based on the taint graph shown as Figure \ref{fig:overview-taint-graph} and the runtime taint data collected (Figure \ref{fig:overview-target-data}) to report the taint behavior.  
We leave the details about taint graph construction and taint propagation to \S\ref{sec:taint-analysis}. 
Note that trace generation, transferring \& decoding, and taint propagation are executed as a pipeline, which is not demonstrated in the toy example.  


%% file: online-selective-tracing.tex
\subsection{Static Identification Analysis}\label{sec:identification-analysis}

\tool requires both dynamic control-flow and data information to replay the taint propagation precisely. 
As for control-flow information, we do not need to record the complete sequence of basic blocks executed. Only the branch decisions are adequate to recover the original execution flow based on the static control flow graph. 
Apart from the control-flow branches, dynamic taint tracking also involves intensive propagation between memory and registers. 
If specific memory addresses cannot be identified, serious precision loss occurs.
To identify the exact memory addresses, one naive idea is to trace all the memory addresses during execution, and then recover the complete execution states during taint propagation.
Such naive design inevitably suffers from high overhead and data loss, as aforementioned. 
We can adopt the state-of-the-art static analysis-based approach such as SelectiveTaint \cite{chen2021selectivetaint} to prune away the irrelevant branches and registers to taint tracking. 
Unfortunately, its effectiveness is marginal -- there are still around 70\% of them required to be recorded dynamically in our experimental study (see \S\ref{sec:optimizations}). This is because 1) the static analysis adopted is conservative; and more importantly 2) branches and registers actually involved in taint tracking are still too intensive.
In fact, not all instructions or memory addresses involved in taint analysis must be recorded at runtime. 
The values of many registers can be statically deduced from that of others based on the intrinsic static code information.
Therefore, we can identify a small set of such necessary registers, and only record their values at runtime. All the others are prevented from tracing since their values can be statically computed.
To this end, we device a dedicated static identification analysis to pinpoint such minimal set of branches and registers essentially to be traced at runtime. 
In particular, the analysis mainly consists of the following three parts.

\MyPara{Redundant Registers Elimination.} 
To obtain the memory addresses required for taint tracking, we need to record the values of registers representing memory addresses. For example, the register \emph{ebx} in \emph{<mov eax, [ebx]>}  has to be recorded.
However, simply tracing all the memory registers in the binary yields significant redundancies. 
In fact, a number of value-flow relations among registers are statically determined. Based on such relation, the values of many registers can be directly deduced from a small set of registers.   
The key is how to identify such a minimal set of essential registers. 

At the high level, Algorithm \ref{alg:target-registers} demonstrates how we determine the minimal set of target registers. Given the value-flow graph (VFG) $\mathcal{V}$, we find all the weakly connected components (WCCs) on it (Line \ref{alg:find-wccs}).
Next, for each WCC (Line \ref{alg:traverse-wccs}), we check whether there is an outgoing address edge in it (Line \ref{alg:has-address-edges}).
If so, we find the root register node of the WCC in the topological order as the target (Line \ref{alg:find-nodes}), and add it to the results (Line \ref{alg:add-nodes}).
Specifically, we first build a VFG on binary to acquire the value-flow relationships among registers. 
A node in the VFG represents either a register or a memory position in an instruction. 
There are two types of edges in the VFG: address edges and value edges. 
Address edges appear between register nodes and memory nodes. An address edge from node $x_r$ to node $y_m$  (\ie, $x_r \stackrel{d}{\dashrightarrow} y_m$) means that the value in $x_r$ is used to calculate the memory address in $y_m$.
A value edge from node $x_r$ to $y_r$ ($x_r \stackrel{v}{\rightarrow} y_r$) means that the value of $x_r$ flows to $y_r$.
 

\begin{algorithm}
\DontPrintSemicolon
\small
    \caption{Register identification}
    \label{alg:target-registers}
    \KwData{A value flow graph $\mathcal{V}$}
    \KwResult{A set of target registers $\mathcal{S}$ }
    \BlankLine
    
    $\mathcal{S} \gets \{\}$\;
    $WCCs \gets $ \textsc{FindWCCs}($\mathcal{V}$)\;\label{alg:find-wccs}
    
    \ForEach{$w \in WCCs$} {\label{alg:traverse-wccs}
        \If{\textsc{HasAddressEdge}($w$)}{\label{alg:has-address-edges}
            $n \gets$ \textsc{FindRootRegNode}($w$)\;\label{alg:find-nodes}
            $\mathcal{S} \gets \mathcal{S} \cup \{n \}$\label{alg:add-nodes}
        }
    }

\end{algorithm}

\begin{figure}[tb!]
	\centering
        \includegraphics[scale=0.42]{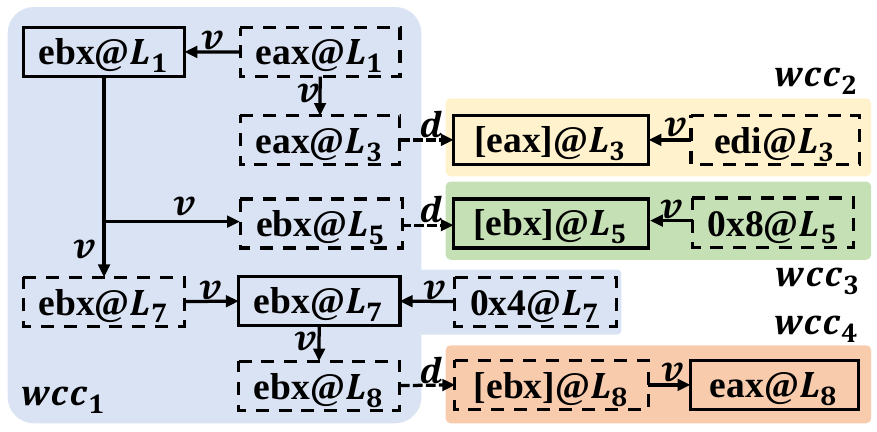}
	\caption{Value flow graph of the binary code in Figure \ref{fig:overview-original-binary}. \label{fig:overview-example-vfg}}
\end{figure}

For example, if we trace the memory addresses in Figure \ref{fig:overview-selectivetatint-traced-target-info}, three registers $eax@L_{3}$, $ebx@L_{5}$, and $ebx@L_{8}$ have to be recorded.
However, after constructing the VFG shown as Figure \ref{fig:overview-example-vfg}, we find that $eax@L_{3}$, $ebx@L_{5}$, and $ebx@L_{8}$ belong to the same WCC (\emph{i.e.}, $wcc_1$). They can actually be derived by $eax@L_{1}$ statically.
Therefore, we only need to trace $eax@L_{1}$ and eliminate $eax@L_{3}$, $ebx@L_{5}$ and $ebx@L_{8}$.
If a WCC has multiple outgoing edges originating from distinct registers, we will trace each register to avoid missing any values.

To further enhance the effectiveness of pruning, we take advantage of alias analysis to enrich the edges of VFG.
In particular, a value edge can be added between two nodes if their registers are must-aliases. In this way, we are likely to discover more and larger WCCs in VFG, thus removing more redundancies. 
We propose a must value-set analysis (MVSA) to generate the must-alias relations among registers. 
Note that here we adopt a conservative must analysis rather than may analysis. Although a may-alias analysis is probably able to find more aliased registers, it also reports false positives -- two registers which are not aliases are determined wrongly as aliases.
Such false positives cause our identification analysis to over-delete some necessary registers, leading to imprecision of taint tracking.  
Note that for the sake of scalability, both VFG construction and MVSA are intra-procedural. 


MVSA follows the lattice of constant propagation \cite{callahan1986interprocedural}, where $\top$ denotes an abstract location (\emph{a-loc}) which can hold any value, 
$i$ represents an \emph{a-loc} holding an exact known value, and $\bot$ represents an \emph{a-loc} whose value is invalid. 
The transfer function of MVSA is a value calculation according to the semantics of each instruction. 
For merging, the value of \emph{a-loc} is set as known \emph{i} only if its value from all predecessor nodes are identical (\ie, value \emph{i}). Otherwise, the value of the \emph{a-loc} is set to invalid.
Note that when a function call is encountered, we conservatively set the value of \emph{a-loc} affected by the function call to invalid.
MVSA terminates once there is no more new \emph{a-loc} added, and each \emph{a-loc}'s value is not changed.
When the analysis is completed, these \emph{a-locs} with the same known value are regarded as must-aliases. We add the corresponding value-flow edges in the VFG.
Note that to ensure the correctness of taint tracking, we explicitly trace all the registers whose must-alias relation cannot be determined, ensuring that no taint data is missed.


\MyPara{Redundant Blocks and Functions Elimination.} 
Apart from registers, some block targets are also redundant.  
As such, we propose a weighted dominator-based algorithm to determine a minimal set of target blocks to be recorded.
Specifically, we first identify the blocks that do not change the taint status.
We consider the taint status (tainted or untainted) of all the registers and memory addresses involved. 
For each input taint status ($\mathcal{IN}_i$) at the entry of a basic block, we calculate the output taint status ($\mathcal{OUT}_i$) at the end of the basic block. If $\mathcal{IN}_i == \mathcal{OUT}_i$ for all the statuses, the basic block is labeled as a taint-unchanged block.
These taint-unchanged blocks are then removed from the CFG. The associated control flow edges are relinked to obtain a reduced CFG.
Next, a dominator graph (DG for short) is constructed based on the reduced CFG, where nodes are CFG nodes, and the edges indicate dominance, post-dominance, or both. Moreover, each node is assigned a weight according to the number of nodes it is dominated.
Subsequently, the SCCDAG (Strongly Connected Component DAG) is built for the DG. Note that nodes in a SCC can dominate each other. Therefore, only one of them needs to be recorded; the execution of other nodes can be statically inferred.
As nodes with higher weights tend to be leaf nodes whose number of times executed should not exceed that of their parent nodes with smaller weights, we choose the node with the largest weight in each SCC as the target block. 

Moreover, we propose a lightweight function-level elimination strategy. 
If a function has neither registers, blocks, global variables involved affecting taint status nor user-defined callees, it does not need to be traced at all. 


\MyPara{Redundant Loop Iterations Elimination.} 
It is well-known that a considerable portion of program execution time lies in the execution of loops. 
If blocks and registers within the loop need to be traced, it undoubtedly leads to high runtime overhead. 
Inspired by loop-invariant code motion \cite{hoe1986compilers} and loop peeling \cite{lin1992compiler} in compiler optimizations, here we eliminate the number of loop iterations to be traced, such that certain blocks and registers in a loop only need to be traced once.
As a result, it reduces not only the overhead at runtime, but also the amount of trace data.

\begin{figure}[h!]
    \centering
    \subfloat[non-once]{\label{fig:non-once-loop}\includegraphics[scale=0.4]{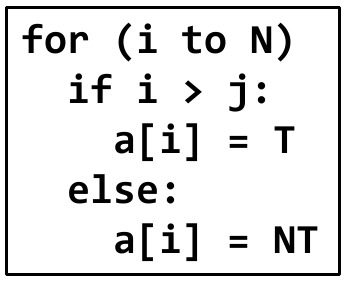}} 
    \hspace{1.0em}
    \subfloat[reg-once]{\label{fig:reg-once-loop}\includegraphics[scale=0.4]{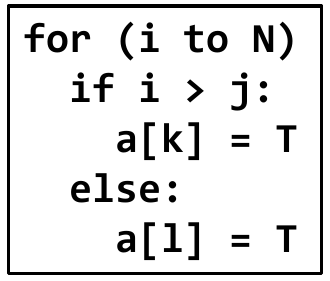}}   
    \hspace{1.0em}
    \subfloat[block-once]{\label{fig:br-once-loop}\includegraphics[scale=0.4]{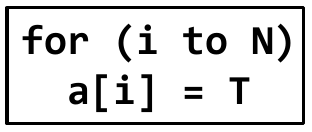}}
    \hspace{1.0em}
    \subfloat[full-once]{\label{fig:full-once-loop}\includegraphics[scale=0.4]{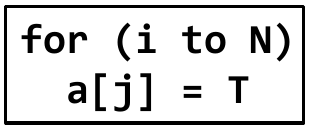}}
    \caption{Example of four types of \emph{once-loop}. \label{fig:once-loop-types}}
\end{figure}

Specifically, we introduce the concept of \emph{once-loop} for taint tracking, \ie, the loop which only needs to be traced for one iteration instead of all iterations. 
The rationale is that not all of the tainted cells (a cell could be a register or a memory address) in the loop are data- or control-dependent on the loop's induction variables (\emph{IVs}). Here for control dependence, we do not take into account the loop branch itself. 
In other words,  the taint status of some registers or memory addresses does not vary between one iteration and multiple iterations. 
Particularly,  a tainted cell in the loop directly or indirectly computed by the \emph{IVs} is called the \emph{data-dependent tainted cell}.
Similarly, \emph{control-dependent tainted cell} indicates the tainted cell that is control-dominated by the \emph{IVs}. 
According to the existence of \emph{data-} and \emph{control-dependent tainted cell}, we classify loops into four types, namely \emph{non-once loop}, \emph{block-once loop}, \emph{register-once loop}, and \emph{full-once loop}.

Figure \ref{fig:once-loop-types} lists four code examples of \emph{once-loop}, where \emph{T} represents a tainted variable, \emph{NT} represents a non-tainted variable, \emph{a} is an array, \emph{j}, \emph{k}, \emph{l} are integer variables defined outside the loop.
In Figure \ref{fig:non-once-loop}, the tainted memory address \emph{a[i]} is both data- and control-dependent on the loop induction variable \emph{i}. For the precise taint tracking,  the  address $a[i]$, branch conditional $i > j$ as well as the loop branch $i \leq N$  have to be recorded at each loop iteration. Therefore, it corresponds to a \emph{non-once loop}.
Figure \ref{fig:reg-once-loop} shows an example of \emph{register-once loop} where the tainted memory addresses \emph{a[k]} and \emph{a[l]} are data-independent but control-dependent on \emph{i}. In this case, the values of \emph{a[k]} and \emph{a[l]} only need to be traced once; while the branch conditional $i > j$ and $i \leq N$ have to be traced at each iteration.
Similarly, Figure \ref{fig:br-once-loop} gives a toy example of \emph{block-once loop}. 
The tainted memory address \emph{a[i]} is data-dependent on \emph{i}. In this case, the block of loop branch $i \leq N$ only needs to be traced once so as to determine whether the loop is executed. The registers inside still need to be traced at multiple iterations.
In Figure \ref{fig:full-once-loop}, the tainted memory address \emph{a[j]} is neither data- nor control-dependent on \emph{i} except for the loop branch.
Therefore, both \emph{a[j]} and $i \leq N$  only need to be traced once.
    


\begin{algorithm}
\DontPrintSemicolon
\small
    \caption{Loop type determination}
    \label{alg:type-of-loop}
    \KwData{a loop $\mathcal{L}$ and a corresponding VFG $\mathcal{V}$ and a CFG $\mathcal{C}$}
    \KwResult{the loop type}
    \BlankLine
    
    $f_d \gets false$ ~  $f_c \gets false$\;
    $IVs \gets $ \textsc{FindLoopIndVars}($\mathcal{L}, \mathcal{V}, \mathcal{C}$)\;\label{alg:find-indv}
    
    \ForEach{taint cell $c \in \mathcal{L}$'s body} {\label{alg:traverse-loop-body}
        \lIf{\textsc{HasReachableEdges}($\mathcal{V}, c, IVs$)}{\label{alg:has-vfg-edge}
            $f_d \gets true$
        }
        \lIf{\textsc{IsDominatedByIVs}($\mathcal{C}, c, IVs$)}{\label{alg:is-dominated}
            $f_c \gets true$
        }
    }
    
    \lIf{$f_d \& f_c$}{\Return  \textbf{non-once} \label{alg:non-once-loop}}
    \lIf{$f_d \& \neg f_c$}{\Return  \textbf{bl-once} \label{alg:br-once-loop}}
    \lIf{$\neg f_d \& f_c$}{\Return  \textbf{reg-once} \label{alg:reg-once-loop} }
    \lIf{$\neg f_d \& \neg f_c$}{\Return \textbf{full-once} \label{alg:full-once-loop}}
    

\end{algorithm}

Algorithm \ref{alg:type-of-loop} lists the pseudo-code to determine the loop type. 
Suppose there is a loop $\mathcal{L}$ and its corresponding value flow graph $\mathcal{V}$ and control flow graph $\mathcal{C}$.
We first find the loop $IVs$ (Line \ref{alg:find-indv}). Then for each tainted cell $c$ in the loop body (Line \ref{alg:traverse-loop-body}), we set the data-dependent flag $f_d$ to true if there exists at least one reachable value-flow edge from $IVs$ to $c$ (Line \ref{alg:has-vfg-edge}) and set the control-dependent flag $f_c$ to true if the loop \emph{IV} control dominates the taint cell (Line \ref{alg:is-dominated}). 
If $f_d$ and $f_c$ are both true (Line \ref{alg:non-once-loop}), it is a \emph{non-once loop}.
If $f_d$ is true and $f_c$ is false (Line \ref{alg:br-once-loop}), it is a  \emph{bl-once loop}.
If $f_d$ is false while $f_c$ is true (Line \ref{alg:reg-once-loop}), it is a \emph{reg-once loop}.
If both $f_d$ and $f_c$ are false, a \emph{full-once loop} is returned (Line \ref{alg:full-once-loop}).


\subsection{Selective Binary Rewriting}\label{sec:binary-rewriter}

Given a minimum of trace points (\ie, target blocks and registers) identified by static analysis (\S \ref{sec:identification-analysis}), we would like Intel PT to selectively trace only these necessary points. 
To this end, we utilize the \emph{ptwrite} instruction to achieve selective tracing.
In particular, for each register determined by the identification analysis, we directly insert right after the original host instruction a \emph{ptwrite} instruction  with the operand being the specific register name. 
For example, to record the register value $eax@L_1$, we insert an instruction \emph{<ptwrite eax>}  after \emph{<mov ebx, eax>}. In other words, the pseudo-code \emph{<record eax>} in Figure \ref{fig:overview-rewritten-binary} is replaced by a physical instruction \emph{<ptwrite eax>}.
For each branch target block to be traced, we insert a \emph{ptwrite} instruction with a dummy operand at the entry of the target block. Whenever a \emph{ptwrite} is executed, Intel PT automatically generates an FUP packet providing the instruction address of the \emph{ptwrite}. In this way, we can determine which block is executed at runtime. 
Furthermore, we consolidate the \emph{ptwrite} instructions for recording target blocks where registers are recorded as well. Specifically, if a register to be traced happens to reside in a target block, we can save the tracing of target block (\ie, a \emph{ptwrite} instruction with a dummy operand) and exploit the trace of register to infer the execution of block.  

For \emph{once-loop}, we insert the \emph{ptwrite} instructions in a predecessor block of the loop header.
If there are more than one predecessor, we create a dummy block and let all the original predecessors jump to the dummy block which then jumps to the loop header.

%% file: offline-taint-graph-parallel-analysis.tex
\subsection{Lightweight Parallel Decoding}\label{sec:decoder}
For the sake of efficiency, the hardware traces generated by PT are stored in a highly compressed format.
A decoding phase is necessary to recover the runtime information from the compressed trace data. 
However, the off-the-shelf decoders (\emph{e.g.}, libipt \cite{libipt-link}, simple-pt \cite{simple-pt-link}) only support sequential decoding, which usually takes a substantial amount of time for processing the tremendous trace data. 
To accelerate the processing, thus lower the analysis latency, we propose a scalable and lightweight parallel decoder specific to taint-related trace data.  
We observed that the PSB packets are generated periodically (\eg, every 4K bytes) for identifying packet boundaries in trace data which can be utilized as split points. The trace data separated by split points can be decoded independently.
As such, we divide the trace data into multiple independent segments, and decode them in parallel.
Moreover, thanks to the selective tracing discussed in \S\ref{sec:binary-rewriter}, all the taint data needed by \tool are recorded into PTW and FUP packets. The decoder thus only needs to care about these two types of packets, while avoiding the heavy decoding along the entire binary. 
Since the instruction addresses and trace points are one-to-one mapped, we are able to recover the branch targets and register values of interest in the original binary.

\subsection{Parallel Taint Analysis}\label{sec:taint-analysis}
For real-time taint tracking, since we do not capture all registers and control flows at runtime, we have to develop a tailored (parallel) taint propagation algorithm based on the taint graph, which combines both the results of static analysis and dynamic tracing information. In this way, we can efficiently conduct static deduction and taint propagation simultaneously.

\MyPara{Static Taint Graph Construction.}
For taint propagation, we introduce the concept of taint graph which includes all the control and data information involved in taint analysis.
A taint graph is intra-procedural and statically determined, where a node corresponds to either a register or a memory location of an instruction.
Besides, certain nodes carry initialization taint information (\eg, taint sources, sinks, and taint flags) based on the taint configuration.
The edges are used to propagate the taint status between two nodes.
They are constructed according to the semantics of instructions and marked by different labels. 
There are five types of edges considered in our implementation, \ie, \emph{a} for value assignment, \emph{d} for memory address propagation, \emph{o} for bit-wise or, \emph{s} for sanitization, and \emph{$bl@L_{i}$} for control-flow propagation.
Figure \ref{fig:overview-taint-graph} shows an example of taint graph, where the node $eax@L_{1}$ is marked as taint source (blue line), and the node $eax@L_{8}$ as sink (red slash). The dashed rectangle indicates a use/from node in taint propagation, while the solid rectangle means a define/to node.
The edge $eax@L_1 \stackrel{a}{\rightarrow} ebx@L_1$ indicates that the taint status of $eax@L_1$ is directly assigned to that of $ebx@L_1$.
For the instruction \emph{<sub ebx, 0x4>}, $ebx@L_7$ is not only a use node but also a define node. 
The edge \dashedbox{$ebx@L_7$} $\stackrel{o}{\rightarrow}$ \solidbox{$ebx@L_7$} denotes that the taint status of \solidbox{$ebx@L_7$} is defined by performing the bit-wise or operation on the taint status of \dashedbox{$ebx@L_7$}.
The edge $ebx@L_8 \stackrel{d}{\dashrightarrow} [ebx]@L_8$ indicates that the taint status of $ebx@L_8$ is assigned to the taint status of the memory address in $[ebx]@L_8$.
The edge $ebx@L_1 \stackrel{bl@L_6}{\dashrightarrow} ebx@L_7$ tells that the taint propagation between $ebx@L_1$ and $ebx@L_7$ is determined by the execution of block target  $L_6$.
Note that the taint graph can support advanced tracking with more diverse edge types.

\MyPara{Sequential Taint Propagation.}
Having the static taint graph and dynamic taint data collected, we start from the taint source, and propagate the taint information along the edges of taint graph accordingly.
Once encountering a branch, we check the target information to figure out along which path the taint propagation continues. 
A branch can also be an entry block of a function, which allows us to achieve inter-procedural taint propagation.
When a memory dereference occurs, we calculate the exact value of memory address according to the register data captured, and check its taint status by looking up a global taint map maintaining all the tainted memory locations.
We repeat the above propagation until all the paths involving taint data collected are traversed. 
Finally, we check the taint status of taint sinks and produce the reports.
For example in Figure \ref{fig:overview-taint-graph}, the node $ebx@L_{1}$ is tainted through the assignment edge from the taint source $eax@L_{1}$.
According to the target block information collected (\ie, $bl@L_{6}:true$), the edge marked as $bl@L_6$ is matched.
The node \dashedbox{$ebx@L_{7}$} is thus tainted.
Next, the node \solidbox{$ebx@L_{7}$} is also tainted according to the \emph{o} edge.
Similarly, the node $ebx@L_{8}$ is also tainted. 
Next, we encounter a dereference edge labeled with \emph{d}. Now, we need to check the taint status of the memory location $ebx@L_8$, whose value is $0x80-0x4=0x7c$. We check it by looking up the global taint map. Assume that the memory location at \emph{0x7c} is tainted, we thus imply that the node $[ebx]@L_8$ is tainted. 
Finally, we deduce that $eax@L_8$ is tainted.

\begin{figure}[htb!]
	\centering
	\includegraphics[scale=0.48]{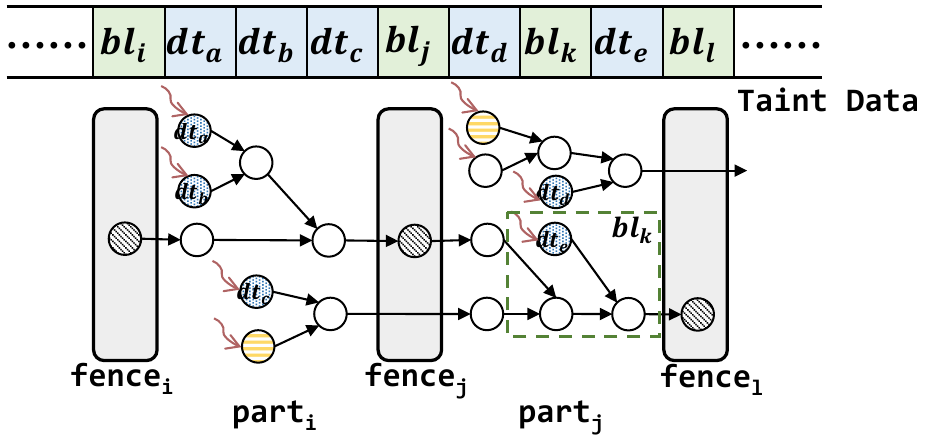}
	\caption{Parallel taint propagation. \label{fig:offline-parallel-taint-analysis}}
\end{figure}

\MyPara{Parallel Taint Propagation.}
To accelerate the propagation, thus minimize the latency between taint triggering and reporting,  we devise a highly parallel propagation algorithm.
Figure \ref{fig:offline-parallel-taint-analysis} demonstrates the skeleton of the parallel taint propagation.
First, we divide the taint graph as well as taint data into multiple partitions (\eg, $part_i$, $part_j$) according to the block target information recorded. 
Within each partition, we assign each thread for each of taint source nodes and entry nodes to start propagation along the taint graph. 
Since taint may propagate across multiple partitions, the status of certain nodes in a partition may not be determined immediately. 
As such, each thread tries its best to propagate as long as possible within the partition, and skips the nodes that are in an unknown status.
These nodes with unknown status need to wait for the taint information provided by its predecessor partition(s).
To this end, we set up a virtual fence between two consecutive partitions to record updates of unknown nodes (\eg, the nodes with black slash in the $fence_i$).
Once the taint information become available later, the propagation continues.  
As the above process is asynchronously parallel, the performance of taint propagation is significantly improved compared to the sequential one, which satisfies the requirement for in-time taint reporting.

%% file: implementation.tex
We use Capstone \cite{anh2014capstone} to disassemble binaries and leverage Angr \cite{shoshitaishvili2016sok} and SelectiveTaint \cite{chen2021selectivetaint} to build sound CFGs. 
The binary rewriting is implemented based on Dyninst \cite{bernat2011anywhere}. 
During hardware tracing, the trace buffer is set to 256MB per CPU core. The model specific registers are set to enable PTW and FUP. 
We directly utilize \emph{mmap} to map trace data onto a buffer in memory, from which the RDMA reads data using the native interface \emph{ibv\_post\_send} provided by libibverbs \cite{libibverbs-link}. At the receiver side, the corresponding trace data is read through \emph{ibv\_post\_recv}.
Our parallel decoder is built on top of libipt \cite{libipt-link}.
For parallel taint propagation, a thread pool with work stealing is  implemented to provide effective scheduling management.

%% file: evaluation.tex
\MyPara{Benchmarks.}
We selected a comprehensive set of well-known and commonly used programs as evaluation subjects covering the Unix utilities (\emph{coreutils}, \emph{gzip}, \emph{scp}, \emph{tar}), network daemons (\emph{exim}, \emph{memcached}, \emph{nginx server}), SPEC CPU INT 2006, and real-world large-scale applications (\emph{PHP}, \emph{MySQL server}).
Due to space limit, we use \emph{coreutils} to represent the 12 subjects from \emph{GNU core utilities} (\emph{i.e.}, \emph{cat, comm, cut, grep, head, nl, od, ptx, shred, tail, truncate, uniq}).
Regarding the test workloads, the workload of Unix utilities is from randomly selected plain-text data; OpenBenchmark \cite{openbenchmark-link} is used for evaluating network applications; SPEC workload is derived from its own test suites; PHP workload is from PHPBench (v1.2.9) \cite{phpbench-link}, and MySQL workload is from its own benchmark suite (\emph{sql-bench}).

\MyPara{Hardware \& Software Environments.}
For trace generation, we used a commodity PC with an Intel i5-12500 6-core CPU with PT and \emph{ptwrite} supported, 16G memory, and 256G SSD, running Ubuntu 22.04.
For decoding and taint propagation, we used a powerful server machine so as to validate the scalability of our parallel taint analysis. In particular, it consists of two Intel Gold 5318Y CPUs, each with 24 cores and 48 threads, 128G memory and 1T SSD, running Ubuntu 22.04. 
Each of the above machines is equipped with a 100Gbps RDMA network card (in particular, NVIDIA ConnectX-6).
We set up RDMA transmission over commodity Ethernet \cite{rocev2-commodity}, where the above two machines are connected by the commodity Ethernet switch. 


Our evaluation answers the following research questions:
\begin{itemize}
    \item Q1: How well does \tool perform? (\S \ref{sec:performance})
    \item Q2: How about the impact of runtime selective tracing and real-time parallel analysis? (\S \ref{sec:optimizations})
    \item Q3: How does \tool perform compared to the state-of-the-art approaches? (\S \ref{sec:comparison})
    \item Q4: How effective is \tool at detecting taint vulnerabilities? (\S \ref{sec:effectiveness})
\end{itemize}

\begin{table}[htb!]
\caption{Performance of HardTaint. Columns \#Inst and \#Func indicate the number of binary instructions and functions, respectively.
Columns AT and RT indicate the time spent in seconds for identification analysis and binary rewriting.
PO, HO, and  TO give the overheads introduced by \emph{ptwrite} execution, hardware tracing, and the total.
Column LC shows the analysis latency in two formats: the absolute time in seconds and the relative. 
}
\centering
\label{tab:online-performance}
\scalebox{0.75}{
\begin{tabular}{ccc|c|c||c|c||c|c|c||l}
\hline
\multicolumn{1}{c|}{Category} & \multicolumn{1}{c|}{Subject}    & Version & \#Inst    & \#Func  & AT (sec.)  & RT (sec.)  & PO      & HO     & TO     & LC (sec./\%)      \\ \hline
\hline
\multicolumn{1}{c|}{\multirow{4}{*}{\begin{tabular}[c]{@{}c@{}}Unix\\ Utilities\end{tabular}}} &
  \multicolumn{1}{c|}{coreutils} &
  8.21 &
  12,214 &
  212 &
  16 &
  3 &
  8.48\% &
  0.38\% &
  8.86\% &
  0.43/5.79\% \\
\multicolumn{1}{c|}{}         & \multicolumn{1}{c|}{gzip}       & 1.3.13  & 15,173  & 220   & 18   & 3  & 6.12\%  & 0.45\% & 6.57\% & 0.56/6.49\% \\
\multicolumn{1}{c|}{}         & \multicolumn{1}{c|}{scp}        & 3.8     & 6,390   & 145   & 9    & 2  & 1.04\%  & 0.05\% & 1.09\%  & 0.65/4.60\%   \\
\multicolumn{1}{c|}{}         & \multicolumn{1}{c|}{tar}        & 1.27.1  & 65,795  & 967   & 118  & 4  & 4.28\%  & 0.48\% & 4.76\%   & 0.54/3.81\%  \\ \hline
\hline
\multicolumn{1}{c|}{\multirow{3}{*}{\begin{tabular}[c]{@{}c@{}}Network\\ Daemons\end{tabular}}} &
  \multicolumn{1}{c|}{exim} &
  4.80 &
  140,847 &
  876 &
  480 &
  11 &
  3.98\% &
  0.56\% &
  4.54\% & 0.38/28.34\% \\
\multicolumn{1}{c|}{}         & \multicolumn{1}{c|}{memcached}  & 1.4.20  & 19,319  & 286   & 25   & 4  & 1.60\%  & 0.77\% & 2.37\%  & 0.26/10.04\% \\
\multicolumn{1}{c|}{}         & \multicolumn{1}{c|}{nginx}      & 1.4.0   & 133,666 & 1,277 & 461  & 8  & 9.09\% & 0.18\% & 9.27\%  & 0.41/6.35\%   \\ \hline
\hline
\multicolumn{1}{c|}{\multirow{12}{*}{SPEC CPU}} &
  \multicolumn{1}{c|}{astar} &
  \multirow{12}{*}{INT 2006} &
  8,802 &
  122 &
  16 &
  6 &
  10.70\% &
  0.39\% &
  11.09\% & 0.55/0.62\% \\
\multicolumn{1}{c|}{}         & \multicolumn{1}{c|}{bzip2}      &         & 11,921  & 103   & 44   & 3  & 4.84\% & 1.07\% & 5.91\% & 0.39/1.83\%  \\
\multicolumn{1}{c|}{}         & \multicolumn{1}{c|}{gcc}      &         & 623,943  & 4,722   & 5,066   & 21  & 11.67\% & 0.37\% & 12.04\% & 0.84/1.01\% \\
\multicolumn{1}{c|}{}         & \multicolumn{1}{c|}{gobmk}      &         & 162,732 & 2,592 & 517  & 7  & 20.23\% & 0.33\% & 20.56\%  & 0.35/0.83\% \\
\multicolumn{1}{c|}{}         & \multicolumn{1}{c|}{h264ref}    &         & 103,512 & 580   & 588  & 5  & 16.13\% & 0.49\% & 16.62\% & 0.53/0.80\% \\
\multicolumn{1}{c|}{}         & \multicolumn{1}{c|}{hmmer}      &         & 58,442  & 572   & 116  & 4  & 1.89\%  & 0.51\% & 2.40\%  & 0.46/0.91\% \\
\multicolumn{1}{c|}{}         & \multicolumn{1}{c|}{libquantum} &         & 9,590   & 138   & 13   & 3  & 1.61\%  & 0.58\% & 2.19\%  & 0.41/2.94\% \\
\multicolumn{1}{c|}{}         & \multicolumn{1}{c|}{mcf}        &         & 2,568   & 54    & 8    & 2  & 6.62\%  & 0.47\% & 7.09\% & 0.46/0.90\% \\
\multicolumn{1}{c|}{}         & \multicolumn{1}{c|}{omnetpp}        &         & 113,389   & 2,141    & 773    & 16  & 10.38\%  & 1.71\% & 12.09\% & 0.57/1.21\% \\
\multicolumn{1}{c|}{}         & \multicolumn{1}{c|}{perlbench}        &         & 219,927   & 1,848   & 968    & 10 & 7.14\%  & 0.86\% & 8.00\% &  0.61/1.73\% \\
\multicolumn{1}{c|}{}         & \multicolumn{1}{c|}{sjeng}      &         & 22,306  & 184   & 27   & 3  & 19.07\% & 0.83\% & 19.90\% & 0.49/0.60\% \\ 
\multicolumn{1}{c|}{}         & \multicolumn{1}{c|}{xalan}        &         & 667,721   & 13,617    & 6,166    & 26  & 8.93\%  & 0.63\% & 9.56\% & 0.76/1.42\%\\ \hline
\hline
\multicolumn{1}{c|}{\multirow{2}{*}{\begin{tabular}[c]{@{}c@{}}Real-world\\ Applications\end{tabular}}} &
  \multicolumn{1}{c|}{PHP} &
  7.4.33 &
  848,252 &
  8,530 &
  18,909 &
  38 &
  10.80\% &
  0.74\% &
  11.54\% & 1.25/3.98\% \\
  \multicolumn{1}{c|}{}         &  \multicolumn{1}{c|}{MySQL} &
  5.6.51 &
  1,857,646 &
  19,426 &
  27,061 &
  57 &
  13.18\% &
  0.55\% &
  13.73\% & 1.43/12.76\% \\ \hline
  \hline
\multicolumn{3}{c|}{Average}                                              & 243,055 & 2,690 & 2,923 & 11  & 8.47 \% & 0.59 \% & 9.06\% & 0.58/4.61\% \\ \hline
\end{tabular}
}
\end{table}

\subsection{Performance}\label{sec:performance}
We run each subject program ten times and collect the average execution elapsed time used.
We configure the taint analysis as the default setting in libdft\footnote{We also run the experiments on libdft which underperforms other state-of-the-art approaches, \eg, SelectiveTaint. The detailed results regarding libdft can be found in our supplementary material.} \cite{kemerlis2012libdft} for a fair comparison with other reference tools (\ie, SelectiveTaint \cite{chen2021selectivetaint}, StraightTaint \cite{ming2016straighttaint}, and TaintPipe \cite{ming2015taintpipe}).
Please refer to the supplementary material for the detailed configurations. 
%
Table \ref{tab:online-performance} lists the performance results for three steps: the time cost of static identification analysis and binary rewriting, the runtime overhead, and the latency of taint propagation. 

\MyPara{Runtime Overhead.}
The runtime overhead of \tool comes from two sources: \emph{ptwrite} execution, and hardware tracing by PT. 
To obtain the exact overhead contributed by each separate part, we run the rewritten binary in two modes: one with PT disabled and one with PT enabled. By disabling PT, no trace data is generated by hardware, where the overhead corresponding to the column PO in Table \ref{tab:online-performance} is only from executing the extra \emph{ptwrite} instructions. 
The column TO indicates the total overhead when \emph{ptwrite} instructions are executed and hardware tracing is enabled. 
TO minus PO results in the value of HO, \ie, the overhead by hardware tracing.
On average, the total runtime overhead introduced by \tool is around 9.06\%.
The overheads on most subjects are less than 10\%, which to some extent is suitable for production-run environments.
In general, the overhead introduced by executing the extra \emph{ptwrite} instructions  dominates the total overhead. 
It varies significantly from different subjects (\eg, 20.23\% for \emph{gobmk}, while only 1.04\% for \emph{scp}).
For \emph{gobmk}, the large PO comes from frequent files reading that causes a large number of registers to be recorded for taint analysis.
For \emph{sjeng}, as plenty of recursive functions are included, the identification analysis fails to optimize redundant blocks/functions. 
Since the traces generated are directly transferred via RDMA in HardTaint, the expensive I/O is bypassed. 
As a result, the overhead in \tool caused by hardware tracing does not exceed 1\% on most subjects, being only 0.59\% on average.
We have reasons to believe that if we propose more effective static analysis to further shrink the set of trace points needed, we can achieve even lower runtime overhead. 

\MyPara{Performance of Static Analysis and Binary Rewriting.}
In Table \ref{tab:online-performance}, columns AT and RT indicate the time spent in seconds for static identification analysis and binary rewriting, respectively.
As can be seen, both static identification analysis and selective binary rewriting are efficient. 
The identification analysis takes about four minutes on average for moderate-sized programs whose number of functions is below the average (\ie, 2,690 in our experiments). 
Importantly, the static analysis can successfully analyze \emph{gcc}, \emph{xalan}, \emph{PHP}, and \emph{MySQL}, validating its good scalability to large-scale real-world programs.  
This is because the key components of identification analysis: VFG construction, MVSA, and the block/function/loop optimizations are intra-procedural and lightweight. 
For large-scale programs, \eg, \emph{gcc}, \emph{xalan}, \emph{PHP}, and \emph{MySQL}, the static analysis still may take several hours (around 1.5 - 7 hours).
Fortunately, it rarely matters since the analysis is a one-time pre-process.
Moreover, the analysis can easily be parallelized using straightforward task parallelism -- performing the same analysis on multiple functions simultaneously. It can thus considerably cut down the analysis time if it is really a concern.
The binary rewriting is even more efficient; most of subjects can be processed in a few seconds. 

\begin{figure}[htb!]
	\centering
	\includegraphics[scale=0.31]
 {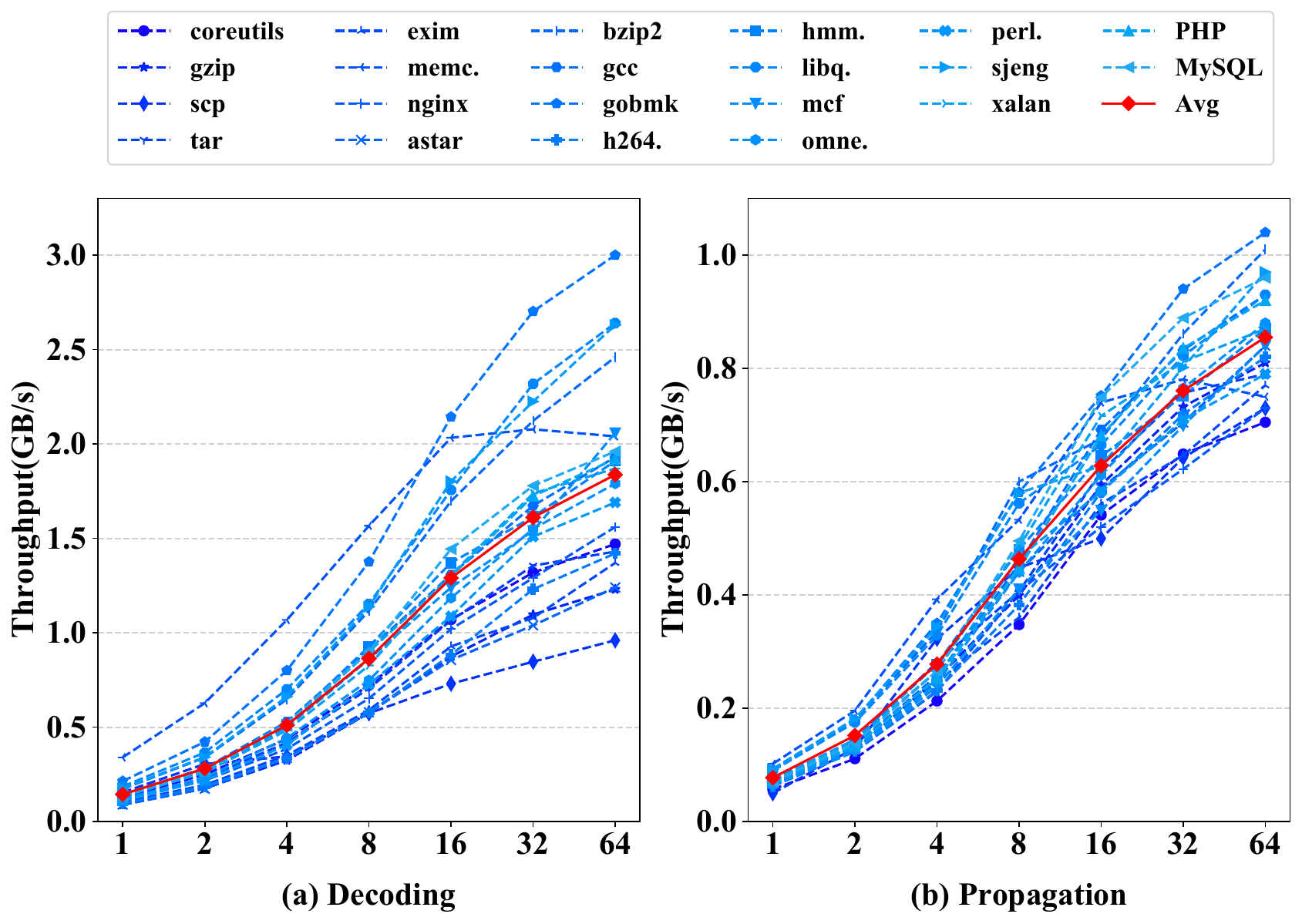}
          \vspace{-1em}
	\caption{Throughput of decoding and propagation under 1, 2, 4, 8, 16, 32, and 64 threads. \label{fig:decode-propagation-scability}}
     \vspace{-1em}
\end{figure}

\MyPara{Throughput of Parallel Decoding and Taint Propagation.}
To better understand how efficient and scalable the taint analysis of \tool is, we evaluate the throughput (\ie, the amount of trace data that can be processed per second) of decoding and taint propagation under 1, 2, 4, 8, 16, 32, and 64 threads.
Figure \ref{fig:decode-propagation-scability} illustrates the scalability of HardTaint's decoding and taint propagation.
The average throughput gradually increases with the growth of threads available. 
With 64 threads, the average throughput of decoding and taint propagation reaches 1.87GB/s and 0.85GB/s, respectively, which is beyond or close to the average speed ($\sim$1.06GB/s) of trace generation in our evaluations.
Thus, we have reasons to believe that \tool can achieve low  latency and in-time reporting.

\MyPara{Analysis Latency.}
To demonstrate the timeliness of \tool in taint reporting, we measure the analysis latency, \ie, the time period from the end of trace generation to the end of taint propagation (shown as Figure \ref{fig:pipeline}). 
Column LC in Table \ref{tab:online-performance} shows the absolute latency in seconds as well as the relative (\ie, the absolute latency divided by program's execution time or trace generation time). 
On average, the absolute latency is 0.58s and the relative latency is only about 4.61\% under 64 threads.
For \emph{exim}, the highest relative latency is due to its relative short execution time to its large-size taint graph that requires more processing time. The same reason is for \emph{MySQL}.
In short, thanks to analysis offloading, pipelining and high-throughput parallel analysis, \tool offers sufficiently low latency for timely taint reporting.

\begin{table*}[thb!]
\caption{Runtime overhead, statistics of instruction counts, and data loss of HardTaint, the representative software-based approach (\ie, SelectiveTaint), and naive hardware tracing (NHT for short). TO indicates the time overhead. $S_{inst}$ denotes the number of instructions to be traced statically. \% in parentheses denotes the proportion of statically traced instructions to the total number of instructions in each binary. $D_{inst}$ denotes the exact number of instructions to be traced dynamically. LT denotes the number of times data loss happened during tracing. EF indicates the effect of lost data on taint propagation, where the larger the proportion of fill within the circle the more severe the data loss. Note that no data loss occurs in HardTaint.}
\label{tab:compare-to-st-naively-tracing}
\centering
\scalebox{0.66}{
\begin{tabular}{c|clc|clc|clcc}
\hline
Subject    & \multicolumn{3}{c|}{HardTaint}  & \multicolumn{3}{c|}{Software-based Approach (\ie, SelectiveTaint)}         & \multicolumn{4}{c}{Naive Hardware Tracing (NHT)}                        \\ \cline{2-11} 
 & TO & \multicolumn{1}{c}{$S_{inst}$(\%)} & $D_{inst}$(x$10^{8}$) & TO & \multicolumn{1}{c}{$S_{inst}$(\%)} & $D_{inst}$(x$10^{8}$) & TO & \multicolumn{1}{c}{$S_{inst}$(\%)} & LT & EF \\ \hline \hline
coreutils  & 8.86\%  & 1,900(15.56)  & 1.66  & 160.75\% & 8,266(67.67)            & 4.59 & 72.07\%  & 8,266(67.67)    & 1,382  & \pie{360} \\ \cline{1-1}
gzip       & 6.57\%  & 2,463(16.23)  & 2.47 & 126.55\% & 10,076(66.41)           & 3.42 & 82.74\%  & 10,076(66.41)    & 1,092  & \pie{360} \\ \cline{1-1}
scp        & 1.09\%  & 950(14.87)    & 0.33  & 17.46\%  & 4,238(66.32)            & 0.65  & 14.54\%  & 4,238(66.32)     & 0      & \pie{0}   \\ \cline{1-1}
tar        & 4.76\%  & 11,379(17.29) & 0.21  & 5.68\%   & 45,630(69.35)           & 1.80  & 68.16\%  & 45,630(69.35)    & 46     & \pie{180} \\ \hline \hline
exim       & 4.54\%  & 12,439(8.83)  & 0.29  & 80.14\%  & 93,058(66.07)           & 0.79  & 87.34\%  & 93,058(66.07)   & 651    & \pie{360} \\ \cline{1-1}
memcached  & 2.37\%  & 1,364(7.06)   & 0.79  & 71.09\%  & 13,676(70.79)           & 2.09  & 13.94\%  & 13,676(70.79)    & 10     & \pie{180} \\ \cline{1-1}
nginx      & 9.27\%  & 16,556(12.39) & 2.51  & 12.83\%   & 92,041(68.86)           & 4.97  & 67.88\%  & 92,041(68.86)   & 255    & \pie{360} \\ \hline \hline
astar      & 11.09\% & 827(9.40)     & 4.02  & 433.71\%  & 6,221(70.68)            & 13.52 & 67.42\%  & 6,221(70.68)     & 1,406  & \pie{360} \\ \cline{1-1}
bzip2      & 5.91\%  & 1,195(10.02)  & 2.07  & 676.96\% & 8,898(74.64)            & 4.38 & 122.28\% & 8,898(74.64)    & 798    & \pie{360} \\ \cline{1-1}
gcc        & 12.04\% & 91,994(14.74) & 6.17  & -         & 432,076(69.25)          & -     & 119.23\% & 432,076(69.25)   & 3,439  & \pie{360} \\ \cline{1-1}
gobmk      & 20.56\% & 18,087(11.11) & 1.09  & 1083.23\% & 118,843(73.03)          & 3.11 & 74.78\%  & 118,843(73.03)   & 537    & \pie{360} \\ \cline{1-1}
h264ref    & 16.62\% & 9,461(9.14)   & 19.17 & 1773.49\% & 78,833(76.16)           & 31.19 & 71.61\%  & 78,833(76.16)    & 731    & \pie{360} \\ \cline{1-1}
hmmer      & 2.40\%  & 5,629(9.63)   & 5.15 & 1572.15\% & 40,989(70.14)           & 7.55 & 102.82\% & 40,989(70.14)    & 1,906  & \pie{360} \\ \cline{1-1}
libquantum & 2.19\%  & 771(8.04)     & 5.37  & 843.27\% & 6,015(62.72)            & 7.82  & 69.93\%  & 6,015(62.72)     & 264    & \pie{360} \\ \cline{1-1}
mcf        & 7.09\%  & 315(12.27)    & 7.06 & 28.84\%   & 1,773(69.04)            & 11.52  & 55.48\%  & 1,773(69.04)     & 627    & \pie{360} \\ \cline{1-1}
omnetpp    & 12.09\% & 13,162(11.61) & 1.77  & 1112.90\% & 91,102(80.34)           & 7.54 & 69.24\%  & 91,102(80.34)   & 1,154  & \pie{360} \\ \cline{1-1}
perlbench  & 8.00\%  & 38,056(17.30) & 2.06  & 305.62\% & 151,012(68.66)          & 5.02  & 58.42\%  & 151,012(68.66)   & 2,088  & \pie{360} \\ \cline{1-1}
sjeng      & 19.90\% & 1,995(8.94)   & 21.91  & 1001.51\% & 15,149(67.91)           & 62.59 & 53.04\%  & 15,149(67.91)    & 233    & \pie{360} \\ \cline{1-1}
xalan      & 9.56\%  & 93,797(14.05) & 5.71  & -         & 527,323(78.97)          & -     & 151.06\% & 527,323(78.97)   & 1,455  & \pie{360} \\ \hline \hline
PHP        & 11.54\% & 89,024(10.49) & 3.77  & -         & \multicolumn{1}{c}{OOT} & -     & 167.59\% & 848,252(100.00)   & 9,656  & \pie{360} \\ \cline{1-1}
MySQL      & 13.73\% & 150,103(8.08) & 15.65 & -         & \multicolumn{1}{c}{OOT} & -     & 83.84\%  & 1,857,646(100.00) & 28,238 & \pie{360} \\ \hline \hline
Average    & 9.06\%  & 26,736(11.00) & 5.19  & 547.42\% & 91,854(70.37)           & 10.15 & 79.69\%  & 211,958(87.21)   & 2,665  & -         \\ \hline
\end{tabular}
}
\end{table*}

\subsection{Impacts of Runtime Selective Tracing and Real-time Parallel Analysis}\label{sec:optimizations}

\input{impact-new.tex}
\begin{figure}[htb!]
     \vspace{-.5em}
	\centering
	\includegraphics[scale=0.32]{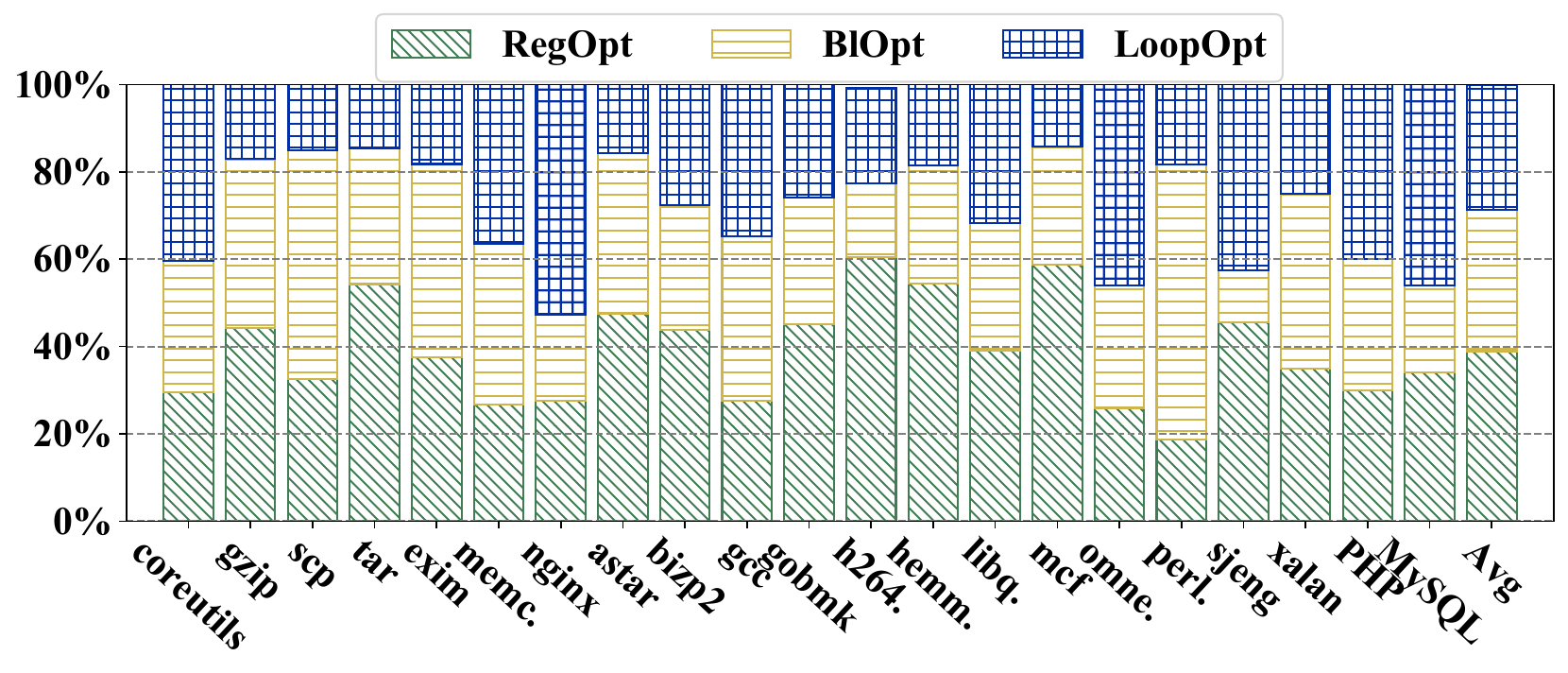}
         \vspace{-0.8em}
	\caption{Percentage of runtime performance improvement for each optimization of static identification analysis. \label{fig:step-by-step}}
 \vspace{-1em}
\end{figure}

\MyPara{Step-by-Step Performance Improvement of Identification Analysis.}
To further understand the impact of each optimization in identification analysis on HardTaint's overhead, we conduct an ablation study. We restrict in turn the application of three optimizations in static identification analysis (\ie, redundant registers elimination (RegOpt), redundant blocks/functions elimination (BlOpt), and loop optimization (LoopOpt)).
Specifically, we generate a version of rewritten binary by applying only RegOpt.
The second version is generated after applying both RegOpt and BlOpt.
The third version is generated after applying all the above three optimizations, \ie, the version reported in Table \ref{tab:online-performance}.
We run each binary version to obtain the respective performance results.
Figure \ref{fig:step-by-step} shows the percentage of performance improvement from each optimization.
It can be seen that all optimizations gain performance benefits. 
The RegOpt contributes mostly on most of the subjects.
On average, BlOpt and LoopOpt contribute roughly the same amount.

\begin{table}[htb!]
\caption{Relative latency of \tool and the baseline.}\vspace{-0.5em}
\label{tab:impact-latency}
\centering
\scalebox{0.85}{
\begin{tabular}{c|c|c|c|c|c}
\hline
Approach  & Unix utils & Networks & SPEC CPU & Real-world Apps & Avg      \\ \hline \hline
HardTaint & 5.17\%     & 14.91\%  & 1.21\%   & 8.37\%          & 4.61\%   \\ \hline
Baseline  & 291.41\%   & 382.16\% & 526.63\% & 874.18\%        & 518.59\% \\ \hline
\end{tabular}
}
\end{table}

\MyPara{Comparison with Naive Offline Taint Analysis.}
To demonstrate the latency improvement using the combination of offloading, pipelining, and parallel taint propagation, we conducted a comparative experiment with the baseline, which loads the entire trace data into memory, decodes it, and performs the sequential propagation. 
Table \ref{tab:impact-latency} shows that the average latency of baseline is as high as 518.59\%, while that of HardTaint is only 4.61\%. \tool achieves orders of magnitude lower relative latency than the baseline. 
This is because that \tool fully utilizes the pipeline to save a large amount of overlapped time.
Meanwhile, combining parallel acceleration of decoding and propagation not only increases the processing throughput, but also helps to further shorten the overall latency.

\subsection{Comparisons with State-of-the-Art Decoupling Approaches}\label{sec:comparison}

\begin{table}[htb!]
 \vspace{-0.8em}
\caption{Comparisons against StraightTaint and TaintPipe.}
\label{tab:compare-with-st-straighttaint-taint-pipe} 
\centering
\scalebox{0.83}{
\begin{tabular}{c|cc|cc|cc}
\hline
\multirow{2}{*}{Subject} & \multicolumn{2}{c|}{HardTaint}  & \multicolumn{2}{c|}{StraightTaint} & \multicolumn{2}{c}{TaintPipe} \\ \cline{2-7} 
      & TO(\%) & LC(\%)  & TO(\%) & LC(\%)  & TO(\%) & LC(\%) \\ \hline \hline
gzip  & 6.57   & 5.39    & 225.00 & 755.00  & 190.00 & 310.00 \\
scp   & 1.09   & 4.60    & 130.00 & 490.00  & 130.00 & 210.00 \\
tar   & 4.76   & 3.81    & 120.00 & 470.00  & 100.00 & 165.00 \\ \hline \hline
astar & 11.09  & 0.62    & 60.00  & 610.00  & 95.00  & 160.00 \\
bzip2 & 5.91   & 1.83    & 200.00 & 1100.00 & 200.00 & 320.00 \\
gcc   & 12.04  & 1.01      & 380.00 & 1120.00 & 220.00 & 440.00 \\
gobmk & 20.56  & 0.83    & 220.00 & 860.00  & 180.00 & 415.00 \\
h264ref & 16.62  & 0.80    & 400.00 & 1330.00 & 320.00 & 520.00 \\
hmmer & 2.40   & 0.91    & 230.00 & 1270.00 & 280.00 & 340.00 \\
libquantum & 2.19   & 2.94   & 120.00 & 690.00  & 50.00  & 150.00 \\
mcf   & 7.09   & 0.90     & 80.00  & 380.00  & 60.00  & 110.00 \\
omnetpp & 12.09  & 1.21    & 50.00  & 850.00  & 70.00  & 180.00 \\
perlbench & 8.00   & 1.73    & 330.00 & 1180.00 & 300.00 & 460.00 \\
sjeng & 19.90  & 0.60   & 300.00 & 1070.00 & 170.00 & 380.00 \\
xalan & 9.56   & 1.42          & 80.00  & 520.00  & 100.00 & 190.00 \\ \hline \hline
Avg   & 9.32   & 1.91   & 195.00 & 846.33  & 164.33 & 290.00 \\ \hline
\end{tabular}
}
\vspace{-0.5em}
\end{table}

As the representative of decoupling approaches, StraightTaint \cite{ming2016straighttaint} online collects information and offline replays taint tracking. 
TaintPipe \cite{ming2015taintpipe} collects dynamic control-flow profiles for constructing pipeline code and spawns multiple threads to execute symbolic taint analysis in parallel.
As neither StraightTaint nor TaintPipe is publicly available, we only compare with the runtime overhead and relative latency reported in the paper.
Table \ref{tab:compare-with-st-straighttaint-taint-pipe} shows the runtime overhead (TO) and relative latency (LC) of HardTaint, StraightTaint, and TaintPipe.
It can be seen that, HardTaint's runtime overhead and relative latency on all subjects are significantly smaller than that of StraightTaint and TaintPipe. 
In particular, the runtime overhead of \tool is 20.92x and 17.63x lower than that of StraightTaint and TaintPipe on average, respectively.
Besides, the latency of \tool is two orders of magnitude smaller than that of both. 
This is because \tool takes advantage of selective hardware tracing to collect taint information, instead of heavily dumping them at runtime via static/dynamic software instrumentation.
Moreover, thanks to the effective pipelining and highly parallel taint propagation, HardTaint's analysis latency is significantly lower than that of the sequential (and symbolic) taint analysis of StraightTaint (and TaintPipe).


\begin{table}[htb!]
\caption{CVEs used for evaluation; \ding{51} denotes that the vulnerability is successfully detected by HardTaint.}
\label{tab:effectiveness-cves}
\centering
 	\scalebox{0.8}{
\begin{tabular}{l|cc|cc}
\hline
\multirow{2}{*}{Software}        & \multicolumn{2}{c|}{CVE}           & \multirow{2}{*}{HardTaint} \\ \cline{2-3}
                                 & ID             & Type              &                           \\ \hline
                                 \hline
\multirow{2}{*}{nginx-1.4.0}     & CVE-2013-2028  & buffer overflow   & \ding{51}                 \\ \cline{2-4} 
                                 & CVE-2013-4547  & validation bypass & \ding{51}                 \\ \hline
nasm-2.14.02                     & CVE-2019-8343  & double free       & \ding{51}                 \\ \hline
\multirow{2}{*}{mini\_httpd-1.19} & CVE-2009-4490  & validation bypass & \ding{51}                 \\ \cline{2-4} 
                                 & CVE-2009-4491  & information leak  & \ding{51}                 \\ \hline
libtiff-4.0.3                    & CVE-2013-4231  & buffer overflow   & \ding{51}                 \\ \hline
mp3gain-1.5.2                    & CVE-2017-14411 & buffer overflow   & \ding{51}                 \\ \hline
ngiflib-0.4                      & CVE-2018-11575 & buffer overflow   & \ding{51}                 \\ \hline
\end{tabular}
}
\end{table}

\subsection{Effectiveness}\label{sec:effectiveness}
Apart from the performance evaluation, we also empirically validate HardTaint's effectiveness on vulnerability detection. 
We selected 8 CVEs of 6 programs which are commonly used by the related work \cite{chen2021selectivetaint, ming2016straighttaint, ming2015taintpipe}. 
The selected CVEs cover a wide variety of software vulnerabilities such as buffer overflow, validation bypass, double free, and information leak.
For experiments, we configure the taint sources and sinks of analysis according the information provided by the CVEs, and then trigger the vulnerabilities with the given PoC files. 
For instance, for the double free vulnerability in \emph{nasm}, we mark the library function \emph{free} as both the taint source and sink. \tool online records the pointer addresses passed to \emph{free}. 
When encountering a \emph{free} function during taint propagation, \tool checks the taint status of the passed pointer address. If it is tainted, \tool report it as a potential double free. If not, \tool marks the pointer address as tainted.
For the buffer overflow vulnerability in \emph{libtiff}, we mark the function \emph{getc} and \emph{fread} as taint source and sink, respectively.
Based on the data recorded, \tool performs taint propagation and detects the variable status passed to \emph{fread}. If the passed variable is tainted, \tool reports a potential buffer overflow violation. 
As shown in Table \ref{tab:effectiveness-cves}, \tool successfully detects all the 8 vulnerabilities.

%% file: impact-new.tex
\MyPara{Comparison with Software-based Approach.}
To demonstrate the necessity of runtime selective tracing (\ie, static identification analysis + hardware tracing), we compare HardTaint and the representative software-based approach (\ie, SelectiveTaint \cite{chen2021selectivetaint}) in terms of runtime overhead, and statistics of instruction counts to be traced.
SelectiveTaint \cite{chen2021selectivetaint} follows the traditional DTA mechanism to perform dynamic taint tracking via software instrumentation. 
It leverages a may-analysis, in particular value-set analysis, to conservatively identify a set of instructions that may be involved in taint tracking. 
All other instructions that are excluded from the resulting set can be safely eliminated from tracing. 
Since the may-analysis adopted over-approximates the actual taint behavior, a significant number of instructions that are, in fact, not involved in taint tracking may be included.
More importantly, even the instructions indeed involved in taint propagation are still too intensive to be recorded at runtime.
In HardTaint, we device a dedicated fine-grained identification analysis to further prune away redundant instructions from the results of SelectiveTaint. 
Compared with HardTaint,  SelectiveTaint lacks both fine-grained identification analysis and hardware tracing. 

Table \ref{tab:compare-to-st-naively-tracing} gives the detailed results, where TO indicates the time overhead.
Column $S_{inst}$  shows the number of instructions identified by static analysis, along with the proportion relative to the total number of instructions in each binary. \textbf{$D_{inst}$} denotes the exact number of instructions traced dynamically, given that one instruction can be executed multiple times.
As can be seen, the number of traced instructions in \tool is much smaller than that in SelectiveTaint, thanks to the fine-grained identification analysis.
To be precise, in terms of the static number of instructions to be traced  (\ie, $S_{inst}$), SelectiveTaint traces an average of 70.37\% of instructions, while HardTaint only traces 11.00\%. In other words, SelectiveTaint eliminates 29.63\% (100\%-70.37\%) of instructions that are irrelevant to taint analysis, while HardTaint's identification analysis further prunes away 59.37\% (70.37\%-11.00\%) of redundant instructions among them.
Comparing the dynamic number $D_{inst}$, HardTaint traces significantly fewer instructions at runtime than SelectiveTaint (\ie, $5.19 \times 10^8$ vs. $10.15 \times 10^8$).
It is validated by evidence that our static identification analysis offers significant contributions over the existing state-of-the-art static analysis.
Beyond question, the runtime overhead of software-based approach (\ie, SelectiveTaint) is significantly (orders-of-magnitude) higher than that of HardTaint. This is because SelectiveTaint has to maintain the large amount of taint information, and perform the heavy taint operations (such as looking up and updating taint information) at runtime.
In contrast, \tool only needs to trace much less data at runtime via much more efficient hardware tracing module. 

Besides, SelectiveTaint fails to instrument the binaries, \emph{gcc} and \emph{xalan} due to runtime errors. 
OOT means SelectiveTaint fails to complete its static analysis within 120 hours, showing its poor scalability. 
In contrast, our static identification analysis is much more efficient and scalable.  On average, HardTaint's identification analysis is 24.25x faster than SelectiveTaint. The detailed comparison about the static analysis time of HardTaint and SelectiveTaint can be seen in the supplementary material. 
Note that for the subjects which SelectiveTaint fails to analyze, \tool takes the raw binary as input instead of the output of SelectiveTaint. 

\MyPara{Comparison with Naive Hardware Tracing.}
To further understand the effectiveness of static identification analysis, we compare the performance of \tool and and naive hardware tracing (NHT for short) with respect to runtime overhead and data loss.
To be specific, NHT takes the analysis result of SelectiveTaint as input, and dynamically records all the relevant branches and registers identified by SelectiveTaint. In other words, NHT has the advantage of hardware tracing, while without our identification analysis enabled.

Due to the severe data loss in NHT, we cannot accurately count the exact number of dynamically traced instructions.
Instead, we collect the number of times data loss happened during tracing denoted by LT and  the effect of lost data on taint propagation represented by EF in Table \ref{tab:compare-to-st-naively-tracing}, where \pie{0}, \pie{90}, \pie{180} and \pie{360} denote no effect, taint-related memory address loss, memory address recovery failure, and control-flow recovery failure, respectively.


As can be seen, although NHT achieves significant performance improvement over software-based approach ($\sim$547.42\% for SelectiveTaint), its overhead  ($\sim$79.69\% on average) is still high and inadequate to meet production-run requirements. 
Even worse, NHT suffers from frequent and severe data loss, leading to catastrophic effects on taint propagation (shown as Columns LT and EF).
In contrast, \tool reduces the overhead to 9.06\%. More importantly, it has no data loss on any subject, offering completely precise taint analysis as the software-based approaches.

%% file: discussion.tex
\MyPara{Practicality.}
\tool primarily targets high-precision dynamic taint analysis in production environments. 
As mentioned, one suitable production environment is the enterprise's data center where the subject applications running on a production machine generate traces at runtime, and dedicated analysis machines exist to process the traces at real-time \cite{google-profiling,prorace-2017}.
In such environment, runtime monitoring overhead is of more crucial concern than the offline analysis cost.


\MyPara{Utility.}
Given the inherent constraints of the decoupled mechanism, there is an inevitable latency between triggering and reporting, even though \tool strives to minimize this delay. As a result, \tool might not be the best fit for situations demanding immediate reporting, like security-critical scenarios. However, there are plenty of scenarios, including privacy monitoring, information leak detection, and diagnostic tasks, where a slight delay is acceptable. \tool is well-suited for these cases.
Besides, \tool has the capability to handle multi-threaded programs. Thanks to Intel PT's support for tracing multi-threaded programs and static analysis to identify shared variables, \tool can accurately restore interleaving information among threads during taint propagation.

\MyPara{Shared Library.}
For commonly used and well-known system libraries (\eg, libc), \tool follows prior works \cite{chen2021selectivetaint, wang2017ramblr, zhu2011tainteraser, sang2023airtaint} by using function summaries to capture the taint propagation relations among the actual arguments, 
In this way, we avoid the heavy online recording of the internals of library functions.
For example, in the case of \emph{memcpy}, its function summary tells that there is always a taint relation from the source to destination. 
Therefore, whenever a call of \emph{memcpy} is encountered, \tool directly assigns the taint status from source to destination without tracking the internals.
For other third-party shared libraries, \tool treats them as the application code, analyzes and traces their complete internal code for taint propagation.


%% file: related-work.tex
\MyPara{Dynamic Taint Analysis.}
Over the past decades, numerous attempts have been made to lower the runtime overhead of DTA.  
Several studies leverage program analysis to remove unnecessary taint tracking.
TaintEraser \cite{zhu2011tainteraser} and TaintStream \cite{yang2021taintstream} adopt function summary to accelerate DTA.  
\cite{jee2012general} proposes Taint Flow Algebra to eliminate redundant taint tracking.
Iodine \cite{banerjee2019iodine} uses an optimistic analysis to elide the proven-safe taint monitoring.
SelectiveTaint \cite{chen2021selectivetaint} adopts value-set analysis \cite{balakrishnan2005wysinwyx} to remove the tracking of taint-irrelevant instructions.
Unfortunately, their effectiveness is tremendously limited by the intrinsic imprecision of program analyses, leading to marginal improvement.
AirTaint \cite{sang2023airtaint}  is a quite recent work which abstracts the basic block-level taint rules to avoid the intensive instruction-level taint logic. We do not select it as the representative software-based approach for comparison since its code is still unavailable (404 error) at the submission time. But, as reported in their paper, the average overhead of AirTaint still easily reaches 60\%, which is much larger than that of HardTaint. 
A group of work accelerates taint propagation with hardware assistance.
The SSE registers in Minemu \cite{bosman2011minemu}, LAHF/SAHF instructions in LIFT \cite{qin2006lift}, and GPUs in FlowMatrix \cite{ji2022flowmatrix} are explored to speed up taint propagation. 
Despite the performance is improved to some extent, their overheads are still far from satisfactory due to the holistic restriction where the heavy taint tracking is tightly fused with the original execution. 
Some work rely on dedicated accelerators \cite{pilato2018tainthls}, co-processors \cite{kannan2009decoupling}, or log-based architectures \cite{chen2006log, chen2008flexible} to accelerate DTA. They are impractical for deployment due to the specific hardware requirement.
While HardTaint can be applied on commodity PCs and supports rich taint analysis without the need for extensive hardware redesign.
Some works \cite{qin2006lift,davanian2019decaf++,galea2020taint} propose fast paths.
Basically, two versions of code are generated -- one with instrumentation for tracking and another without instrumentation.
During execution, the transition data is maintained to determine which path to take.  
However, this maintenance still leads to the considerable overhead.
Another category is to decouple the taint reasoning from the original execution, thus offloading the taint analysis on other processes/threads \cite{ming2015taintpipe,cui2015practical,quinn2016jetstream,ruwase2008parallelizing,nightingale2008parallelizing,jee2013shadowreplica,ming2016straighttaint, patil1995efficient}.
Some approaches \cite{patil1995efficient,nightingale2008parallelizing} run an instrumented version simultaneously with the original non-instrumented program in other processes/cores. 
However, as is well known, it is notoriously difficult to reproduce the identical program execution  without imposing too much overhead due to the existence of non-determinism and runtime exceptions.
Other works collect runtime information during the execution of the original program, and then use this information for taint propagation. 
Among them, FlowWalker \cite{cui2015practical} records runtime execution snapshots and offline replays them for taint analysis.
ShadowReplica \cite{jee2013shadowreplica} and TaintPipe \cite{ming2015taintpipe} leverage the profiling information to construct optimized taint code and perform the reconstructed taint tracking using shadow threads in parallel. 
StraightTaint \cite{ming2016straighttaint} collects taint information at runtime via profiling and memory dumping, followed by an offline symbolic taint analysis.
JetStream \cite{quinn2016jetstream} partitions executions into epochs using deterministic record and performs local taint analysis on each core. The results from all cores are aggregated to obtain a complete result.
Unfortunately, these decoupling approaches adopt software-based instrumentation for information logging and dumping, which introduces high overhead and analysis latency. 
In contrast, \tool exploits selective hardware tracing and parallel taint propagation to lower runtime overhead and analysis latency, respectively.

\MyPara{Acceleration via Hardware Tracing.}
Hardware tracing has been widely applied to various areas, such as program security \cite{ge2017griffin, hu2018enforcing, gu2017pt}, debugging \cite{cui2018rept,yagemann2021arcus, zhangissta23armdiagnosis}, testing \cite{chen2019ptrix, schumilo2017kafl, li2022muafl}, etc. 
Despite its advantages, it has not yet been adopted to accelerate DTA. 
The major concern is that DTA requires prohibitively heavy and intensive data and control information, which readily causes non-trivial overhead and severe data loss. 
\tool fills this gap and proves its feasibility.


%% file: conclusion.tex
We propose and implement HardTaint, the first system to the best of our knowledge leveraging modern hardware tracing module to realize production-run DTA on a commodity PC. 
\tool offers ultra-low runtime overhead via static analysis and selective hardware tracing. 
Moreover, in-time taint reporting with low latency is also achieved, benefiting from the analysis offloading, task pipelining, and highly parallel taint analysis.
Evaluations over well-known subjects demonstrate the ultra-low runtime overhead and negligible analysis latency of HardTaint, while possessing the completely precise taint detection capability. 